\documentclass{acmtog}

\usepackage{wrapfig}

\usepackage{amsmath}
\usepackage[utf8x]{inputenc}
\usepackage{color}
\usepackage{wrapfig}
\usepackage{amssymb}
\usepackage[ruled,vlined]{algorithm2e}

\def\BigColSep{\setlength{\arraycolsep}{0pt}}

\acmVolume{VV}
\acmNumber{N}
\acmYear{YYYY}
\acmMonth{Month}
\acmArticleNum{XXX}  
\acmdoi{10.1145/XXXXXXX.YYYYYYY}

\acmVolume{28}
\acmNumber{4}
\acmYear{2009}
\acmMonth{August}
\acmArticleNum{106}  
\acmdoi{10.1145/1559755.1559763}

\begin{document}

\markboth{N. Ray and D. Sokolov}{On Smooth $3D$ Frame Field Design}

\title{On Smooth Frame Field Design} 

\author{N. RAY
\affil{INRIA Lorraine}
\and
D. SOKOLOV
\affil{Université de Lorraine}}




\acmformat{}

\maketitle

\begin{bottomstuff}
Authors' addresses: Nicolas.Ray@inria.fr dmitry.sokolov@loria.fr.
\end{bottomstuff}

\begin{abstract}

We analyze actual methods that generate smooth frame fields both in $2D$ and in $3D$.
We formalize the $2D$ problem by representing frames as functions (as it was done in $3D$),
and show that the derived optimization problem is the one that previous work obtain via ``representation vectors''.
We show (in $2D$) why this non linear optimization problem is easier to solve than directly minimizing the rotation angle of the field,
and observe that the $2D$ algorithm is able to find good fields.

Now, the $2D$ and the $3D$ optimization problems are derived from the same formulation (based on representing frames by functions).
Their energies share some similarities from an optimization point of view (smoothness, local minima, bounds of partial derivatives, etc.),
so we applied the $2D$ resolution mechanism to the $3D$ problem.
Our evaluation of all existing $3D$ methods suggests to initialize the field by this new algorithm, but possibly use another method for further smoothing.




\end{abstract}

\section{Introduction}

In computer graphics, a frame field can be defined on a surface ($2D$) or inside a volume ($3D$).
For each point of the domain, it defines a set of $4$ (resp. $6$) unit vectors invariant by rotations of $\pi/2$ around the surface normal (resp. around any of its member vector).
The main motivation to study these fields is to split the quad and hexahedral remeshing problems into two steps: (1) the design of a smooth frame field, (2) and the partitioning of the domain by quads or hexes aligned with the frame field.
Our objective is to unify the formulation of the $2D$ and $3D$ frame field design problems, and to use it to extend an efficient $2D$ algorithm to the $3D$ case.

\begin{figure*}
\centerline{ \includegraphics[width=\linewidth]{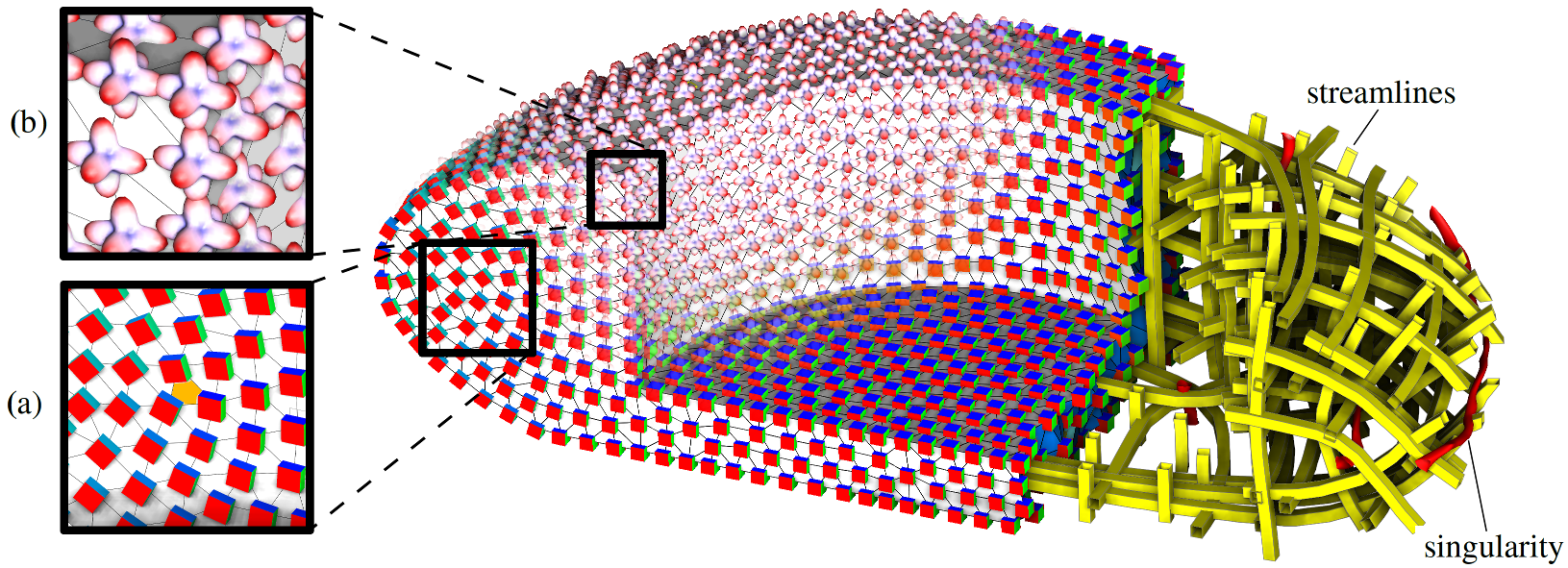}}
\caption{Orthonormal frames (close-up (a)) are represented by spherical harmonic functions (close-up (b)), attached to each vertex of a tetrahedral mesh.
Streamlines and singularities of the field are shown in yellow and red, respectively.
}
\label{fig:teaser}
\end{figure*}

In most cases, frame field design is formalized as the following optimization problem: find the smoothest frame field subject to some constraints. What makes them different from each others is obviously the dimension of the frames ($2D$ or $3D$), but also the definition of the field smoothness, the expression of the constraints, and the optimization method. Interestingly, the $2D$ case and the $3D$ case are addressed by very different strategies:
\begin{itemize}
\item In $2D$, the frame field design problem can be restated as a vector field design problem thanks to the introduction of the {\bf ``representation vector''}.
In local polar coordinates, each vector of a frame has the same angle modulo $\pi/2$, if we multiply it by $4$ we obtain a unique representation vector (modulo $2\pi$).
It is easy to derive optimization algorithms acting on the representation vectors.
For simplicity reasons, we limit ourselves to planar frame fields and use the algorithm proposed by Kowalski \emph{et al.}~\cite{kowalski2012} as reference.
\item In $3D$, it is not possible to extend the idea of ``representation vector''.
Instead, Huang \emph{et al.}~\cite{Huang:2011} propose to represent frames by {\bf functions} defined on the sphere, refer to figure~\ref{fig:teaser} for an illustration.
A definition of the field smoothness is derived from this representation and optimized in a two step procedure:
(1) initialization based on optimization of spherical harmonics coefficients in \cite{Huang:2011} or front propagation of boundary constraints in \cite{Li:2012},
followed by (2) smoothing iterations performed by L-BFGS on Euler angles representation of frames.
\end{itemize}

Thus our goal is to better understand how $2D$ and $3D$ problems are related to each other and to extend \cite{kowalski2012} to $3D$.
We first show that \cite{kowalski2012} can be derived with the formalization approach inherited from the $3D$ case, and then we extend it to $3D$ by the same logical flow.
\textit{Busy readers interested in only reproduction of the method can skip to implementation section~\S\ref{sec:implementation3D}, the only required tool is a linear solver, all calculations are given explicitely.}

\subsection*{The $2D$ algorithm with frames represented by functions }

Solutions developed for $3D$ are very different from $2D$ solutions because the ``representation vector'' trick does not extend nicely into $3D$.
To unify both problems, we propose to go in the other direction \S\ref{sec:star}: we apply the functional frame representation to the $2D$ case.
By doing so, we found another way to introduce the ``representation vectors'': they appear as coefficient vector of the function decomposed in the Fourier function basis \S\ref{sec:2Dframerep}.
Following the logical flow introduced for the $3D$ case, we derive an estimation of the field smoothness \S\ref{sec:2Dobjfct} and formulate the corresponding optimization problem \S\ref{sec:implementation2D}.
We end up with exactly the formulation of $2D$ methods based on the ``representation vectors''.

This common formulation of the $2D$ problem is not the simplest one: the difference between adjacent frames is not evaluated by the rotation between them as in \cite{Bommes:2013:IMR:2461912.2462014,NSDF},
but approximated by the Euclidian distance in a plane.
We observe the impact of this approximation on the objective function \S\ref{sec:toypb}, and show how it simplifies the optimization problem \S\ref{sec:result2D}.

The resolution mechanisms of the $2D$ optimization problem are strongly inspired by the geometric intuition of the representation vector field.
To make abstraction of this intuition, we re-explain the algorithm from~\cite{kowalski2012} with the notations/vocabulary introduced by the functional representation of frames.

\subsection*{Extension of the $2D$ algorithm to $3D$}
Now that we can describe the $2D$ algorithm without referring to the representation vector, it is possible to extend it to $3D$.
Using the notations introduced in the $2D$ case, we describe the $3D$ version in \S\ref{sec:framework3D} and extend the $2D$ optimization mechanism to $3D$ \S\ref{sec:implementation3D}.

A first difficulty was to find the expression of the boundary conditions because the boundary frames are free to rotate around the surface normal \S\ref{sec:alignmentcondition}.
Incomplete enforcing of this condition by Huang \emph{et al.}~\cite{Huang:2011} results in a poor initialization of the optimization procedure as evaluated in \S\ref{sec:ch1eval}.

The second difficulty is that the frame is represented in a $9D$ function basis, but the set of functions that corresponds to a frame has dimension $3$
(the $3D$ rotation that brings the axis aligned frame to the current frame).
The extension of the normalization of the representation vector in \cite{kowalski2012} becomes: find the nearest point on the $3D$ manifold of the set of functions representing a frame.

Our extension of \cite{kowalski2012} nicely completes the tool set of $3D$ frame field design algorithms \S\ref{sec:results}.
Our initial field has lower energy, often a better topology, and is more robust to surface noise.
The optimization step is easy to implement from the initialization step, is competitive when the initial topology is good enough, but does not outperform the current state of the art.

\subsection*{Previous works}

The orientation of objects is commonly represented by symmetric tensors in physics to model the orientation distribution of fibers.
For example, Moakher~\cite{Moakher} introduced the notion of cubic orientation distribution functions.
However, in computer graphics, more compact representations are often preferred: ``representation vectors'' in $2D$ and vectors of spherical harmonics coefficients in $3D$.

\subsubsection*{On surface}

The optimization of a frame field inside a $2D$ shape is very similar to the optimization of direction fields on surfaces \footnote{The differences between these two problems (angle defect, hard constraints, curvature fitting term, etc.) have an impact on the optimization algorithm, as detailed in the supplemental material}.

The problem of direction field design on surfaces was introduced by \cite{Hertzmann00illustratingsmooth} for orienting strokes in non photo realistic rendering. Directions are represented by an angle rotation $\theta$ per vertex, and the smoothing is performed by a non linear solver (BFGS). The "representation vector" was not introduced yet, and the optimization mechanism was similar to actual approach of $3D$ frame fields smoothing. Solving with a representation vector $v = (r\cos(\theta), r\sin(\theta))$ for each $\theta$ was used in \cite{PGP} for faster results, and improved later for better control over the field topology \cite{DFD}. Based on the similar representative vectors,  \cite{Palacios:2007} propose to control the field topology by local operations. For direction fields without constraints, \cite{keenan:2013} proved that the norm of the representative vector does not affects the result, leading to finding optimal direction fields by solving an eigenvector problem.

Directly working with angle $\theta$ allows to perfectly control the field topology, but at the expense of solving a mixed integer system \cite{NSDF,mixtquad}. In \cite{Palacios:2007}, representative vectors are introduced with its duality with $N^{th}$ order traceless symmetric tensors. This relation is very useful to unify $2D$ and $3D$ frame fields.

\subsubsection*{Inside a volume}

The pioneer work of \cite{Huang:2011} discovered that each frame can be represented by a spherical harmonic. Their initialization step defines a smooth spherical harmonic field then, for each sample, defines the frame that better aligns with the spherical harmonic. This initial solution is then improved by rotating each frame. These rotations are defined by Euler angles and obtained by minimizing the field smoothness with L-BFGS and enforcing the boundary alignment by a penalty term. 

Li {\it et al.} \shortcite{Li:2012} propose an alternative initialization method. They optimize a $2D$ frame field on the volume boundary, convert it to a $3D$ frame field by adding the surface normal and its opposite, then propagate it inside the volume. The resulting field is then smoothed by optimizing a rotation for each frame, as in \cite{Huang:2011}, but with an improved optimization scheme. They also optimize the singularity graph of the field by local combinatorial operations, as done by \cite{Tengfei:2014}.

Our extension of \cite{kowalski2012} is optimizing for the same objective function, but with a very different solution mechanism. It performances are compared with previous work in \S\ref{sec:results}.

\setlength\columnsep{.7\marginparsep}
\setlength\intextsep{.7\marginparsep}

\section{Functional representation of frames in $2D$}
\label{sec:star}

This section introduces how to optimize $2D$ frame fields using a functional representation for each frame.
While we do not claim any technical contribution in this section, we think that it is important to reformulate existing concepts
using the functional representation, because $3D$ case inherits exactly the same difficulties and the intuition
we gained in $2D$ helps to motivate the choices made for $3D$ fields.
We derive an energy and the boundary conditions from this new representation.
The resulting optimization problem is exactly the one usually solved by direction field algorithms based on the standard ``representation vector''.
We show on a toy and a real examples that this optimization problem is much easier than directly minimizing the frame rotation.
We then present the algorithm~\cite{kowalski2012} that we extend to $3D$ in the following section.

\subsection{Problem settings}

\begin{figure}[htbp]
\centerline{\includegraphics[width=2.5cm]{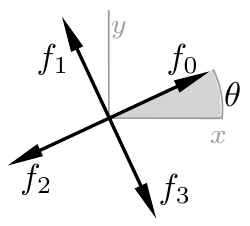}}
\caption{A $2D$ frame is a set $f$ of $4$ vectors $f_0,f_1,f_2,f_3$ invariant under rotation by $\pi/2$. Its angle representation is the rotation $\theta$ between the global axis $x$ and $f_0$.}
  \label{fig:2dframe}
\end{figure}

Given a $2D$ shape, frame field design in $2D$ consists of finding a smooth frame field aligned with the boundary of the shape. We formulate it as minimizing the field curvature, based on the following definitions \footnote{The problem is directly presented in discrete settings. Interesting results on its (non trivial) continuous counterpart are given in supplemental materials.}:

\begin{itemize}
\item A {\bf frame} is a set of $4$ unit vectors $f=\{f_i\}, i \in [0,3]$ that is invariant by a rotation of $\pi/2$ (Figure~\ref{fig:2dframe}). It can be represented by the angle $\theta$ such that $\forall i, f_i = (\cos(\theta+i\pi/2),\sin(\theta+i\pi/2))$.
\item A {\bf frame field} is a frame per vertex of a $2D$ shape triangulation.
\item The {\bf boundary constraint:} a frame located on a boundary vertex must have one of its member vectors equal to the normal on the boundary.
\item The {\bf rotation angle between two frames} is the angle $\Delta\theta$ of the rotation that transforms one frame into the other. This angle being defined modulo $\pi/2$, we choose the $\Delta\theta$ with minimum absolute value.
\item The {\bf curvature of a frame field} is the sum over each edge of the squared rotation angle between frames that are defined at the edge extremities.
\item A triangle is said to be {\bf singular} \footnote{Frame field topology is further discussed in supplemental material.} if the sum of the rotation angles over his boundary is not equal to $0$.
\end{itemize}

Representing frames by angles is simple, but it makes the field curvature hard to optimize~\cite{mixtquad,NSDF}, and it does not extend nicely in $3D$.
For these reasons, we propose an alternative representation based on functions, and use this curvature based formalization as a reference.

\subsection{Functional approach: frame representation}
\label{sec:2Dframerep}

\begin{figure}[t]
\centerline{\includegraphics[width=6cm]{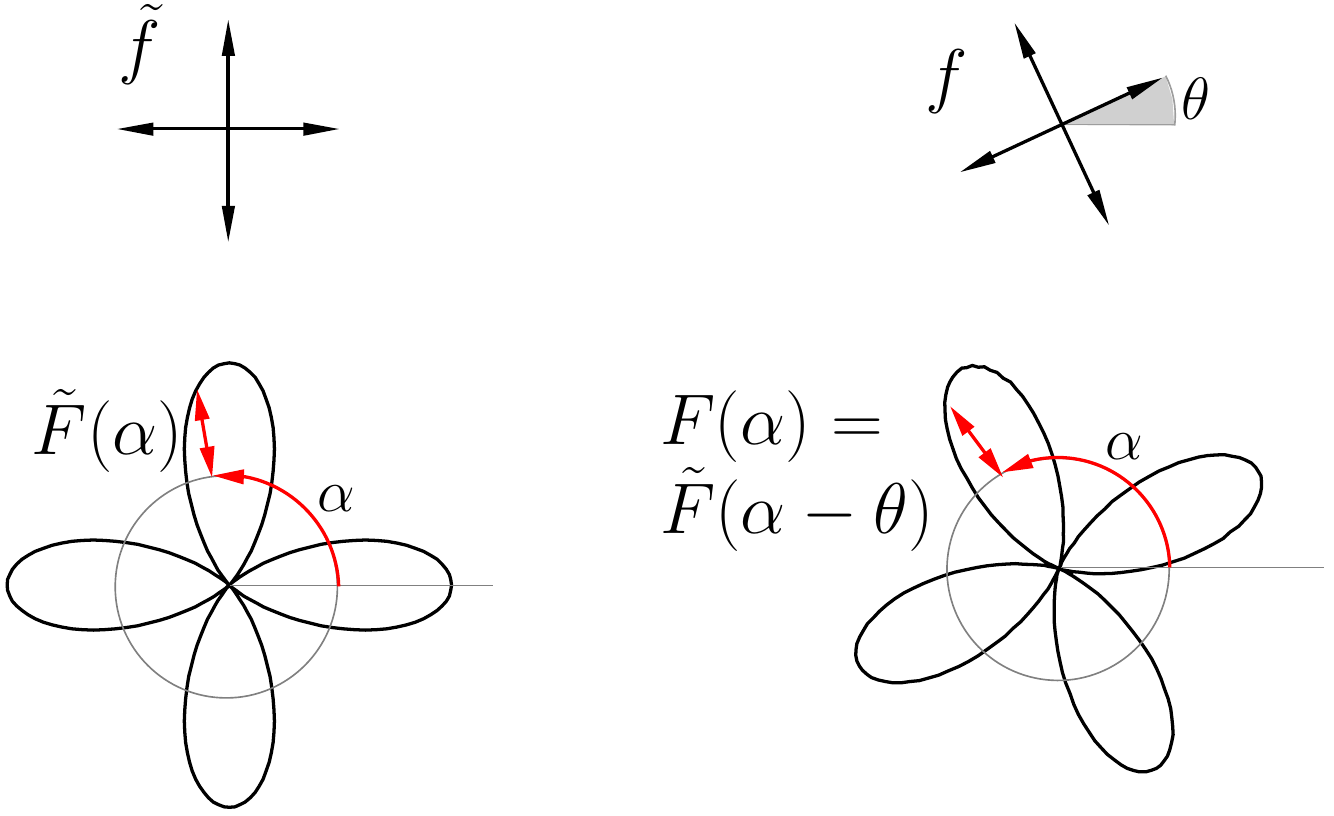}}
\caption{Parametric plot of the reference frame representation $\tilde{F}(\alpha)$ (left) and an arbitrary frame $F(\alpha)$ (right).
The plot is made using $x(\alpha)= (1+F(\alpha))\cos(\alpha)$ and $y(\alpha) = (1+F(\alpha))\sin(\alpha)$ for $\alpha \in [0,2\pi[$.
It is easy to see that corresponding frames are aligned with maxima of the representation functions.}
  \label{fig:2DfunctionRep}
\end{figure}

The frame aligned with coordinate axes is called the reference frame $\tilde{f} = \{(1,0),(0,1),(-1,0),(0,-1)\}$.
Instead of using the angle approach, we represent it by the function $\tilde{F}(\alpha) = \cos(4\alpha)$ with $\alpha \in [0, 2\pi[$ (Figure~\ref{fig:2DfunctionRep}--left), that exhibits the same $\pi/2$ rotation invariance as the frame.

Any other frame $f$ can be obtained by a rotation of $\tilde{f}$ by angle $\theta$.
The functional counterpart is to rotate the graph of the function $\tilde{F}$, namely
any frame $f$ can be represented by a function $F(\alpha) = \tilde{F}(\alpha-\theta) = \cos(4(\alpha-\theta))$ with $\alpha \in [0, 2\pi[$ (Figure~\ref{fig:2DfunctionRep}--right).

A compact representation of these functions is given by the trigonometric relation $f(\alpha) = \cos(4\alpha-4\theta) = \cos(4\theta)\cos(4\alpha) + \sin(4\theta)\sin(4\alpha)$: we see that a frame function $F$ can be represented by a $2D$ vector of coefficients $a = (a_0,a_1)^\top = (\cos(4\theta), \sin(4\theta))^\top $ in Fourier basis $B=(\cos(4\alpha),\sin(4\alpha))$ i.e. $F=Ba$.

A {\bf coefficient vector $a$ is feasible} if and only if there exists $\theta$ such that $a = (\cos(4\theta), \sin(4\theta))^\top$.
Geometrically, $a$ is constrained to live on a curve parameterized by $\theta$.
This curve represents, in coefficient space, all possible rotations of the reference frame.
In $2D$, this curve is the unit circle, so the constraint on $a$ is simply : $a^\top a=1$.

At this point, we can observe that the coefficient vector $a$ is exactly the representative vector used in the direction field literature.
We can also notice that expressed in Cartesian coordinates, our reference frame function $\tilde{F}$ is the polynomial $4(x^4+y^4)-3$ restricted to the unit circle,
thus it is also equivalent to the traceless symmetric $4^{th}$ order tensors manipulated in~\cite{NROSY}.

\subsection{Functional approach: objective function}
\label{sec:2Dobjfct}
We have defined the field curvature as the sum over each edge of the squared difference between frames at the edges extremities. In our formalization, the difference between two frames (at vertices $i$ and $j$) is no longer the rotation angle, but the $L^2$ norm of the difference between the corresponding functions~: $\int_0^{2\pi} (F^j(\alpha) -F^i(\alpha))^2  d\alpha$. It leads to the new objective function:
\begin{eqnarray}
E &=& \sum_{ij} \int_0^{2\pi} (F^j(\alpha) -F^i(\alpha))^2  d\alpha\nonumber\\
& = & \sum_{ij} \int_0^{2\pi} (Ba^j -Ba^i)^2  d\alpha \nonumber\\
& = & \sum_{ij} (a^j -a^i)^\top\left(\int_0^{2\pi} B^\top B d\alpha \right)(a^j -a^i) \nonumber
\end{eqnarray}
As the function basis $B$ is orthogonal, and all functions are of norm $\sqrt{\pi}$, so the expression simplifies to:
\begin{equation}
\label{eq:smoothNRJ2D}
E  =  \pi \sum_{ij} \|a^j -a^i\|^2
\end{equation}

\subsection{Functional approach: constraints}
As discussed in \S\ref{sec:2Dframerep}, the first constraint is clearly that the variables $a^i$ must be feasible (i.e. there exists a frame represented by $a^i$).

The second constraint is that frames of boundary vertices $i$ must have one member vector equals to the normal direction.
If $\theta^i$ denotes the normal direction, the frame can be directly fixed by satisfying two equations:
\begin{eqnarray}
\label{eq:2Dboundarylock}
a^i_0 &=&\cos(4\theta^i) \\
a^i_1 &=&\sin(4\theta^i) \nonumber
\end{eqnarray}

However, as we already have the feasibility constraint ${a^i}^\top a^i=1$, enforcing only one equation has the same effect:
\begin{equation}
\label{eq:2Dboundary}
a^i_0 \cos(4\theta^i)  + a^i_1 \sin(4\theta^i) = 1.
\end{equation}

\subsection{Toy example}
\label{sect:toy}
It is natural to ask the question: \textit{``Does minimizing our energy minimizes the field curvature as well?''}

Two frames $f^i$ and $f^j$ are represented by $a^i$ and $a^j$, both located on the unit circle.
The field curvature measures the circle arc length between them, whereas our $L^2$ norm is the chord length between $a^i$ and $a^j$.

From a practical point of view, we want to produce smooth fields, so most edges will have low curvature. In this case the objective function $E$ is almost proportional to the field curvature. If, however, two adjacent frames are not similar (e.g. they are close to singularities), then the function $E$ is smoother than the field curvature, making it easier for the optimization algorithm to move singularities.

Let us illustrate our intuition on a very simple interpolation example: a chain of four vertices having its extremities locked.
The toy problem is therefore to find two frames interpolating the extremity frames.

\begin{figure}
\centerline{\includegraphics[width=\linewidth]{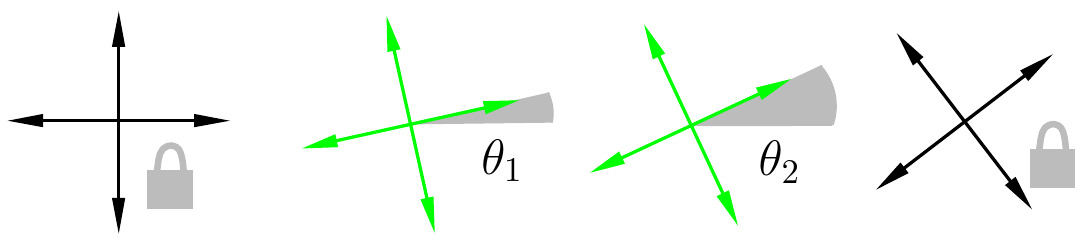}}
\vspace{3mm}
\centerline{\includegraphics[width=\linewidth]{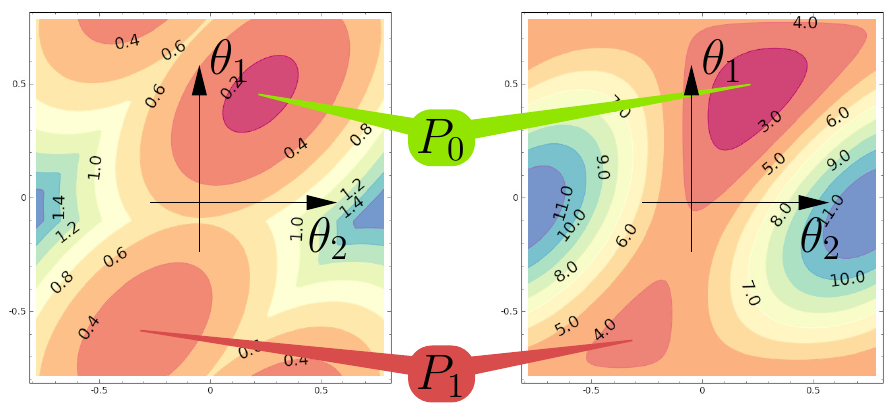}}
\caption{Top row: interpolation problem with two locked extremity frames.
Bottom row, left: field curvature plot. Bottom row, right: our objective function $E$.
The plots are made as functions of the rotations $\theta_1,\theta_2$ of interpolated frames. Both functions share same local minima $P_0$ and $P_1$.}
  \label{fig:nrjplot2D}
\end{figure}

If we represent two free frames by their angle $\theta_1$ and $\theta_2$, we can plot and compare the field curvature versus our objective function $E$ (Figure~\ref{fig:nrjplot2D}). The field curvature is not smooth (it is piecewise quadratic) and we can observe that there are two local minima. Our objective function is smooth, and has exactly the same minima on this example. Note that it could also have a smaller number of minima e.g. if the constrained frames are more similar.

\begin{figure}
\centerline{\includegraphics[width=\linewidth]{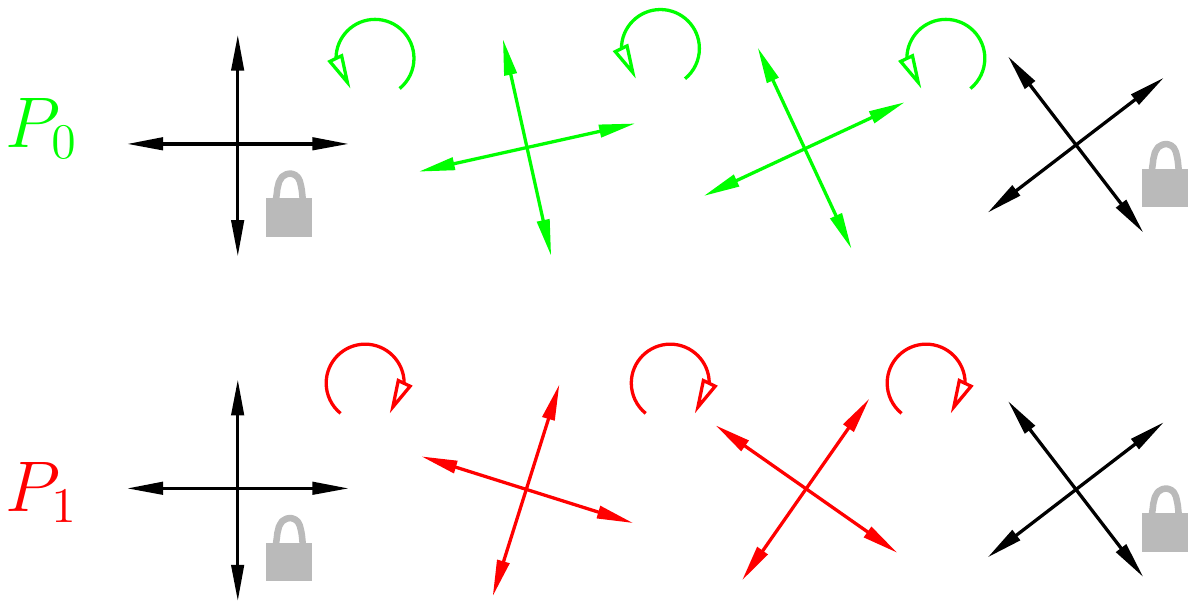}}
\caption{Two minima for the toy problem shown in Figure~\ref{fig:nrjplot2D}. $P_0$ turns the frames counterclockwise while $P_1$ turns clockwise. $P_0$ minimizes energy $E$ and has better field curvature.}
  \label{fig:nrjplot2Db}
\end{figure}

Figure~\ref{fig:nrjplot2Db} shows the frames corresponding to minima $P_0$ and $P_1$: they differ by the sense of rotation. The point $P_0$ minimizes objective function $E$ and has better field curvature. In next two sections we expose current state of the art approach to minimization of the objective function.

\subsection{Implementation}
\label{sec:implementation2D}
We have to minimize our objective function $E$ (eq.~\eqref{eq:smoothNRJ2D}) with linear equality constraints on boundary vertices (eq.~\eqref{eq:2Dboundarylock} or eq.~\eqref{eq:2Dboundary}) and quadratic equality constraints ${a^i}^\top a^i =1$ for each vertex $i$.

We use~\cite{kowalski2012}'s algorithm to solve this problem. It finds an initial solution by relaxing the quadratic feasibility constraints on $a^i$ and finding the nearest feasible solution. Then it performs a number of smoothing iterations to ameliorate the solution. In order to respect the feasibility, the quadratic constraints are linearized at each smoothing step.

\paragraph*{Initialization}
Here, we relax the feasibility constraints, so we need to minimize the quadratic function $E$ subject to linear boundary constraints. To do it, we simply replace the linear constraints by a strong penalty term in the objective function, leading to a new quadratic function to optimize. This penalty method is very simple and sufficient in practice.

More precisely, the new quadratic function is expressed in the form $\|AX-b\|^2$ where $A$ is a matrix, $X$ our variable vector ($X_{2i}=a^i_0$ and $X_{2i+1}=a^i_1$) and $b$ is a vector. The system of equations $AX-b=0$ is constructed line-by-line:
\begin{itemize}
\item {\it initial objective function $E$:} for each edge $ij$, we add two equations (eq.~\eqref{eq:smoothNRJ2D}):
\begin{align*}
\sqrt \pi (X_{2i}-X_{2j})&=0\\
\sqrt \pi (X_{2i+1}-X_{2j+1})&=0
\end{align*}
\item {\it boundary constraints:} for each boundary vertex $i$, we add two equations (eq.~\eqref{eq:2Dboundarylock}):
\begin{align*}
\lambda X_{2i} &=  \lambda \cos 4\theta^i\\
\lambda X_{2i+1} &= \lambda \sin 4\theta^i, 
\end{align*}
where we set $\lambda=100$ in our experiments.
\end{itemize}

From $A$ and $b$, we just need to solve the linear system $A^\top AX=A^\top b$ to obtain a minimizer of $\|AX-b\|^2$.
Then from $X$ we can obtain an initialization of $a^i$ by projecting corresponding vectors on the set of feasible coefficients:
$$
a^i\leftarrow (X_{2i}, X_{2i+1})^\top/\|(X_{2i}, X_{2i+1})\|.
$$

\paragraph*{Smoothing iterations}
Each smoothing iteration is similar to the initialization problem, except that we add to the objective function a new quadratic penalty term that corresponds to a linear approximation of the feasibility constraint as done in~\cite[p. 6]{kowalski2012}. As before it is expressed by a new set of linear equations when constructing $A$ and $b$: for each vertex $i$, we add one equation $\lambda (X_{2i}a^i_0 + X_{2i+1}a^i_1-1) =0$, where $a^i$ denotes the solution obtained at the previous iteration.

To solve linear systems we use OpenNL library~\cite{OpenNL}: it automatically constructs $A^\top AX=A^\top b$ from the set of linear equations and then solves it by the conjugate gradient method.

\subsection{Toy problem revisited}
\label{sec:toypb}

This section explains our optimization approach on the toy problem already presented in \S~\ref{sect:toy}.
As we mentioned before, at the initialization step we relax the constraints of feasibility of $a^i$.
Unfortunately we can not plot the corresponding energy since without the constraints it becomes four-dimensional.

\begin{figure}
\centerline{\includegraphics[width=\linewidth]{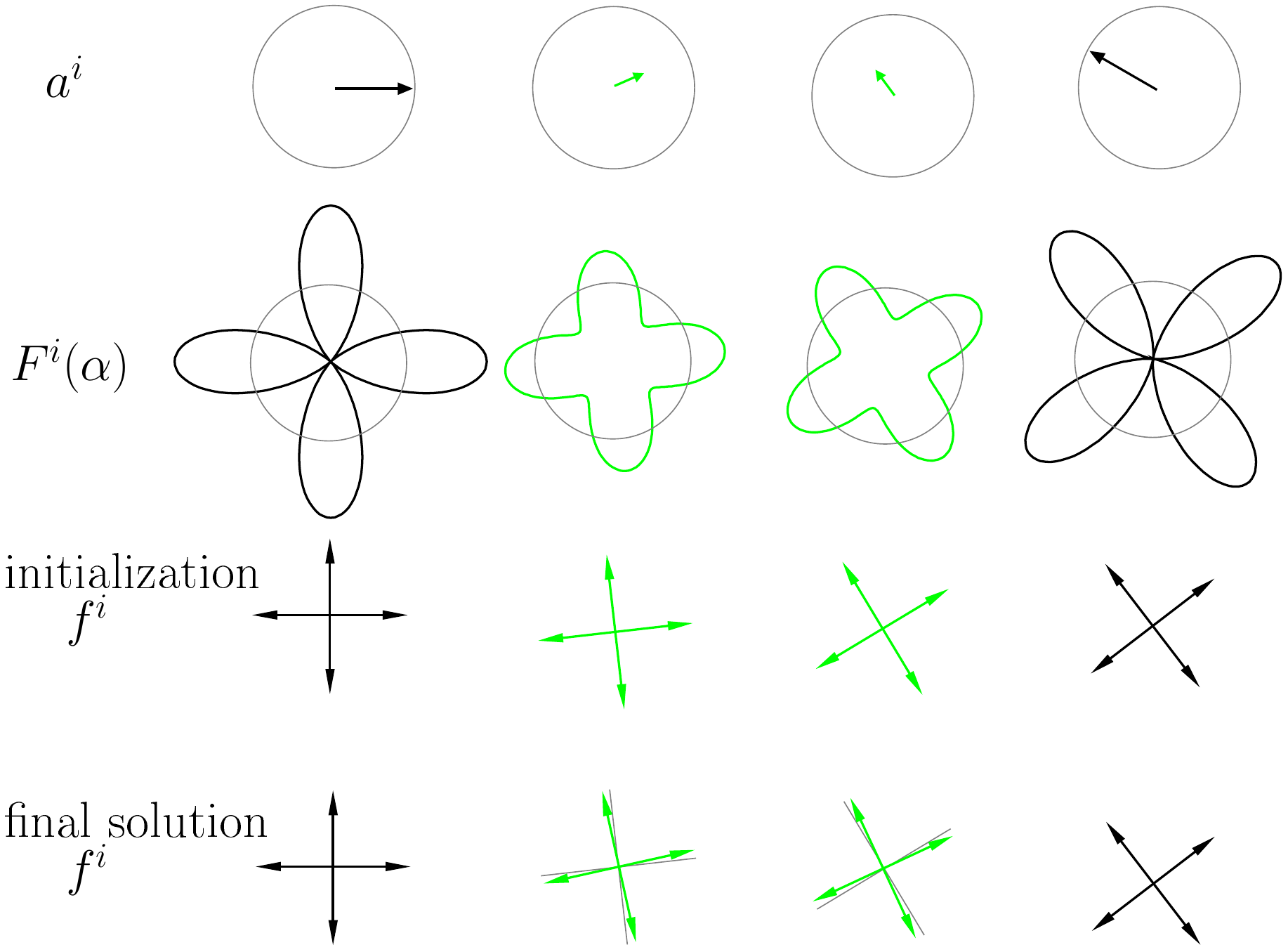}}
\caption{First row: initialization stage solution. Second row: corresponding functional interpretation. Third and fourth rows: frames obtained by the projection of initialization $a^i$ and after smoothing iterations.}
\label{fig:toy_initialization}
\end{figure}

Top row of Figure~\ref{fig:toy_initialization} shows the solution of the initialization stage.
Intuitively, we allow the points $a^i$ not to be on the unit circle.
Hence $a^1$ and $a^2$ lie on the chord between locked points $a^0$ and $a^3$.
Second row of Figure~\ref{fig:toy_initialization} shows the corresponding functions and the third row
gives the frames obtained by projecting points $a^i$ on the circle of constraints.

Note that the initialization stage produces the correct sense of rotation (Figure~\ref{fig:nrjplot2Db}).
However it does not directly produce the optimum point $P_0$.
The reason being that the initialization stage produces points $a^1$ and $a^2$ (before normalization) in the way that
all three chord segments are equal: $\|a_0-a_1\|=\|a_1-a_2\|=\|a_2-a_3\|$.
But after projecting the points onto the feasible circle (3rd row of Figure~\ref{fig:toy_initialization}) the corresponding arc lengths are not equal.
Therefore, we need a few smoothing iterations to reach the optimum (Figure~\ref{fig:nrjplot2Db}---bottom row).

\subsection{Results}
\label{sec:result2D}

\begin{figure*}[!htb]
\centerline{\includegraphics[width=15cm]{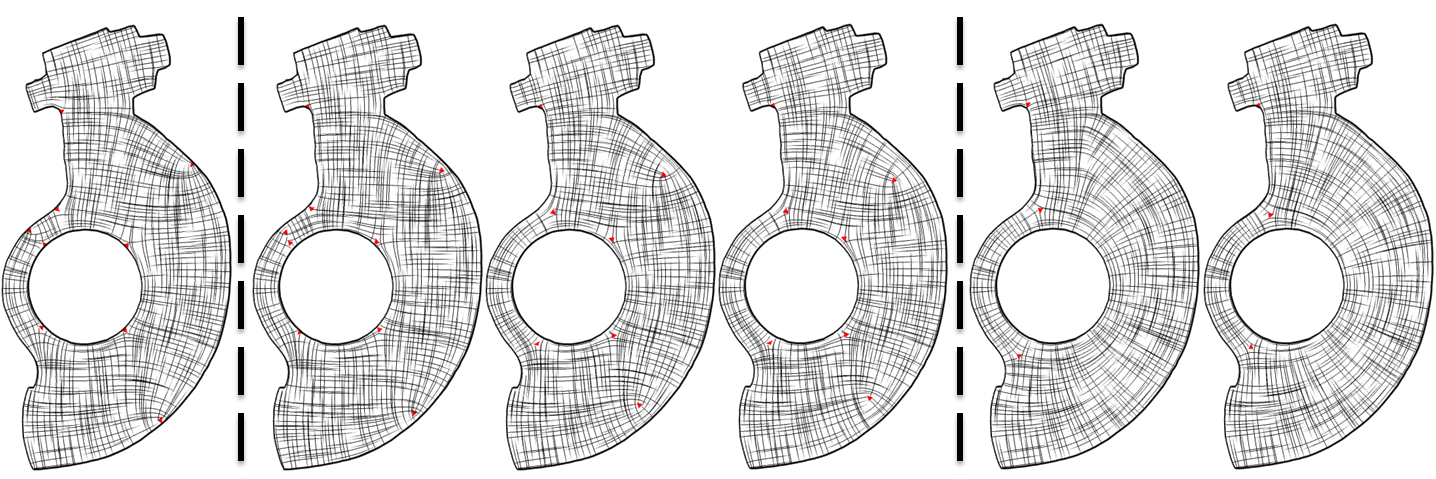}}
\caption{{\it Evaluation of $2D$ frame field optimization algorithms:} singular triangles are highlighted in red. Compared algorithms are: {\bf(left)} steepest descent of the field curvature, initialized with an axis aligned frame field, {\bf (middle)} smoothing iterations with our objective function $E$ initialized with axis aligned frame field after $10^2$, $10^3$ and $10^6$ iterations (from left to right), {\bf(right)} initialization step alone (left) and the initialization step followed by $10^3$ smoothing iterations (right).}
  \label{fig:rockerslice}
\end{figure*}

Figure.~\ref{fig:rockerslice}--left shows the optimization of the field curvature by a gradient descent algorithm, initialized by an axis aligned frame field. 
We obtain a frame field that bends to match the boundary constraints, but its singularities remain on the boundary. 
The system is only able to reach the local minima corresponding to the initial field topology. The field curvature \footnote{The value of the field curvature is relevant only for comparing different fields on the same mesh.} is $34.21$.

Figure~\ref{fig:rockerslice}--middle shows the optimization using only the smoothing iterations, initialized by an axis aligned frame field.
We observe that smoothing iterations were able to slightly move the singularities away from the border and even to merge two singularities.
It results in a better field curvature (equal to $31.41$, $24.01$ and $23.95$ after $10^2$, $10^3$ and $10^6$ iterations).

Figure~\ref{fig:rockerslice}--right shows that the initialization step alone finds a solution with a simple topology and a lower field curvature ($20.84$). Smoothing iterations further decrease the field curvature (to $17.91$).

These observations suggest that our initialization step is mandatory, and smoothing iterations further improve the result. However, it is important to notice that each iteration takes approximately the same time as the initialization step.

\begin{figure}[!htb]
 \centerline{\includegraphics[width=8cm]{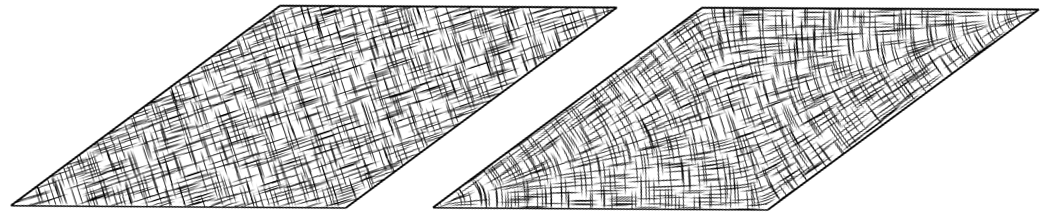}}
\caption{{\it Boundary constraints for global optimization:} if we only use one constraint (eq.\ref{eq:2Dboundary}), the initialization step finds a constant frame field on a parallelogram (left). It is therefore mandatory to lock the frames to get the proper boundary constraints (right).}
  \label{fig:losange}
\end{figure}

\paragraph*{Boundary constraints}
When working only with feasible solutions a single equation (equation~\eqref{eq:2Dboundary}) is sufficient to enforce each boundary constraint.
Using it for the initialization step is wrong: for example, a normal constraint of angle $\theta=0$ forces $a_0=1$ but let $a_1$ free. As illustrated in Figure~\ref{fig:losange}, it can produce very bad frames fields. Therefore we must use two separate equations~\eqref{eq:2Dboundarylock}.

There exists a similar issue in $3D$: Huang et al~\cite{Huang:2011} use a $3D$ boundary condition that is not sufficient for the initialization step. 
It leads to a poor initialization for their smoothing algorithm, making it very slow, and getting possibly locked with a bad initial topology. This issue is discussed in details in \S\ref{sec:results}.

\section{Optimization of $3D$ frame fields}
\label{sec:framework3D}

Our objective is to extend to $3D$ the optimization algorithm presented  in previous section.

In $2D$, our framework allows to retrieve the representation vector that was the key to efficient optimization of direction fields. We now extend our framework to $3D$, find a generalization of this representation vector, express the boundary alignment condition with respect to this representation, and derive the optimization algorithm.

\subsection{Problem settings}

The problem is to define, inside a $3D$ shape, a smooth frame field that is aligned with the boundary of the shape. 
We are working in discrete settings on a tet mesh.
The problem to minimize the field curvature is defined as follows:
\begin{itemize}
\item The {\bf reference frame} $\tilde{f}$ is the set of $6$ unit vectors forming normals of a cube aligned with coordinate axes (Figure~\ref{fig:3d_fct}).
\item A {\bf frame} is the reference frame rotated by a $3\times 3$ matrix $R$: $f = R \tilde{f}$.
Note that multiplying a matrix by a set is a slight abuse of notation, it means that we obtain a new set where each vector is rotated by the given matrix.
\item A {\bf frame field} is the definition of a frame for each vertex of the tet mesh.
\item The {\bf boundary constraint}: The frame of a boundary vertex must have one of its member vectors equal to the normal of the boundary.
\item The {\bf rotation angle between two frames} is the minimal angle of rotation that brings one frame to the other.
\item The {\bf curvature of a frame field} is the sum, over each edge, of the squared rotation angle between adjacent frames.
\item A tet is called {\bf singular} if any of its triangles is singular.\footnote{We can define what a singular tet is, but we are not able to characterize them by an equivalent of the index in $2D$. This fact is discussed in the supplemental material.} The triangle $ijk$ is singular if and only if $R^{ij}\times R^{jk}\times R^{ki} \neq Id$, where $R^{ij}$ denotes the rotation matrix that brings the frame $f^i$ to the frame $f^j$.
\end{itemize}


\subsection{Frame representation}
\label{sec:3Dframerep}

The reference frame $\tilde{f}$ is represented by the function $\tilde{F} = \sqrt{\frac{7}{12}} Y_{4,0}+\sqrt{\frac{5}{12}} Y_{4,4}$, where $Y_{l,m}$ is the real valued spherical harmonic of degree $l$ and order $m$.
These harmonics are sometimes called tesseral~\cite[p. 74]{ferrers1877elementary}. The list of harmonics of degree 4 can be found in~\cite[p. 239]{gorller1996rationalization}).
Function $\tilde{F}$ is defined as $\mathbb{R}^3\rightarrow \mathbb{R}$, but we are interested by its restriction to the unit sphere $\mathbb{S}^2\rightarrow \mathbb{R}$.

\begin{figure}
 \centerline{ \includegraphics[width=6cm]{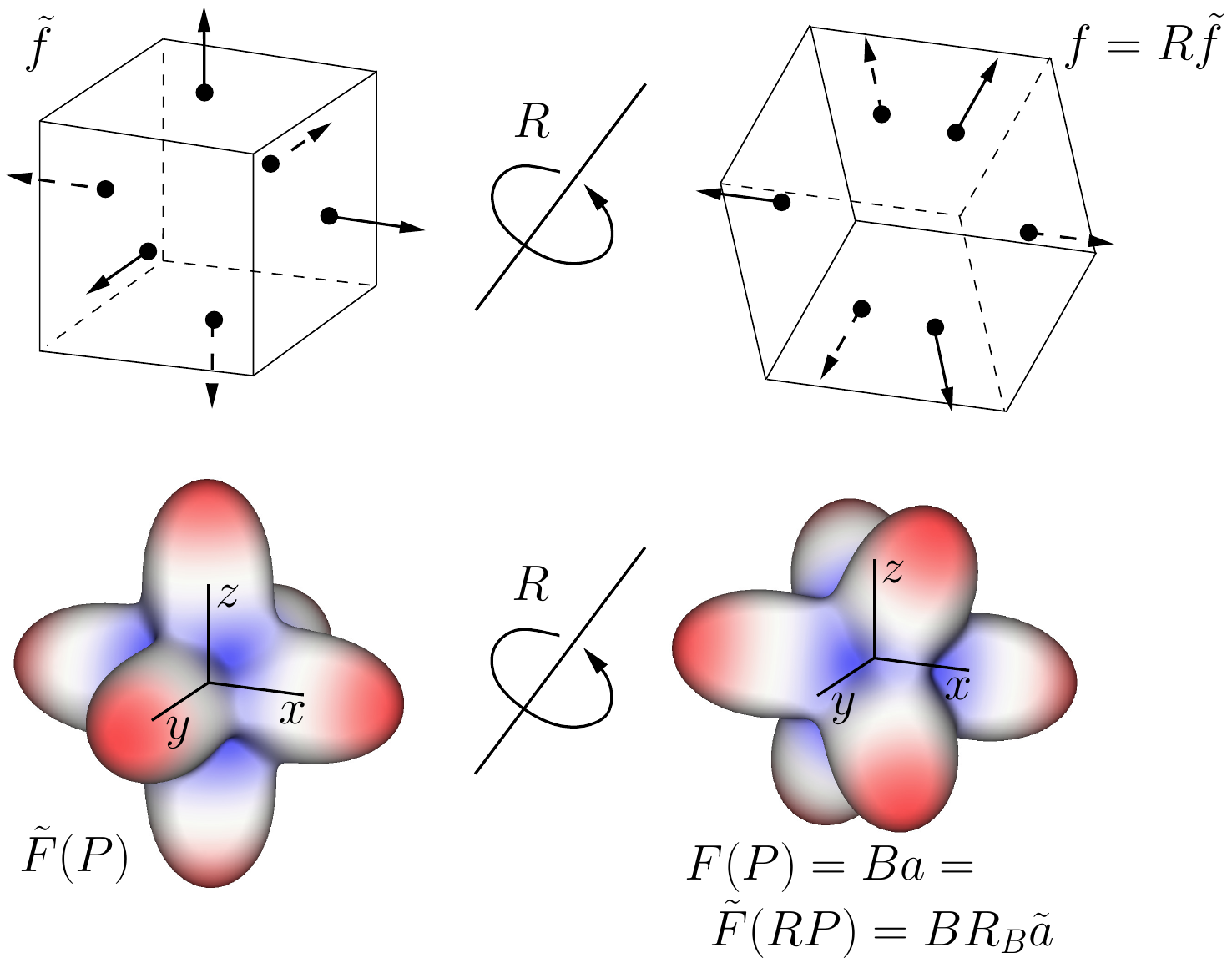}}
\caption{A frame $f$ is the reference frame $\tilde{f}$ rotated by a $3\times 3$ matrix $R$. The plots of the corresponding functions $F$ and $\tilde{F}$ are also rotated by $R$, and their coefficients vectors verify $a=R_B\tilde{a}$ where $R_B$ is a $9 \times 9$ rotation matrix given in \S\ref{sec:9drotation}.}
  \label{fig:3d_fct}
\end{figure}

Any other frame $f$ can be obtained as a rotation of $\tilde{f}$ by a matrix $R$. It is represented by the function $F(P) =  \tilde{F}(RP)$, where $P$ is a point of the unit sphere (Figure~\ref{fig:3d_fct}).

$Y_{l,m}$ forms a functional basis over the unit sphere with an interesting property: applying a rotation to a spherical harmonic of degree $l$ produces another harmonic of degree $l$.
As a consequence, since we represent the reference frame by a sum of two spherical harmonics of degree 4, each frame function $F$ can be represented in the basis $B=(Y_{4,-4},Y_{4,-3},\dots,Y_{4,4})$.
Using it, we can rewrite the expression for the reference frame function as $\tilde{F} = B\tilde{a}$ with  $\tilde{a} = \left(0,0,0,0,\sqrt{\frac{7}{12}},0,0,0,\sqrt{\frac{5}{12}}\right)^\top$.
Any other frame $f = R \tilde{f}$ can be represented by $F=Ba$, where $a=R_B\tilde{a}$ with $R_B$ being a $9\times 9$ rotation matrix acting on coefficients space.
Appendix~\ref{sec:9drotation} describes the construction of the rotation matrices.

A {\bf feasible coefficient vector} $a$ is a vector that can be written as $a = R_B\tilde{a}$ where $R_B$ is a $9D$ rotation matrix that can be derived from a $3D$ rotation.  Geometrically, $a$ is constrained to be on a manifold of dimension $3$ embedded in the $9D$ coefficient space.

At this point we can consider the coefficient vector $a$ as an extension of the representative vector used in the direction field literature. It is also the representation introduced in \cite{Huang:2011}.

\subsection{Objective function}

As in $2D$, the objective function is the sum, over each edge $ij$, of the squared difference between frames located at the edges extremities. In our formalism, the difference between two frames is not the rotation angle, but the $L^2$ 
norm of the difference between the corresponding functions: $\int_0^{2\pi} (F^j(\alpha) -F^i(\alpha))^2  d\alpha$. It gives the energy:
\begin{equation}
E = \sum_{ij} \int_0^{2\pi} (F^j(\alpha) -F^i(\alpha))^2  d\alpha\nonumber
\end{equation}
Here, the function basis $B$ is orthonormal, so the expression simplifies to:
\begin{equation}
\label{eq:smoothNRJ3D}
E  =  \sum_{ij} \|a^j -a^i\|^2
\end{equation}

\subsection{Constraints}
\label{sec:alignmentcondition}

There are two types of constraints: each coefficient vector $a^i$ must be feasible, and boundary frames must have one vector aligned with the normal of the volume boundary. The first constraint is presented in the frame representation section, and will be enforced by our optimizer in a dedicated ``projection'' step (the $3D$ counterpart of the normalization of $a$ in the $2D$ case). 
Here we focus on the boundary constraint.

\subsubsection*{Smooth vertex} We assume first that there is only one normal associated with the vertex, it can be computed as the average normal vector of incident boundary triangles. 

{\bf Case 1:} the normal is equal to the $z$ axis.

Let us first consider the case where the fixed vector (the surface normal) is the axis $z$. If we rotate $\tilde{F}$ around $z$ by angle $\theta$, we obtain $a = R_B\tilde{a}$ with $R_B$ being a rotation around $z$. The simple structure of $R_B$ together with the null coefficients of $\tilde{a}$ gives the equation:
\begin{equation}
a = \left(\sqrt{\frac{5}{12}}\sin 4\theta,0,0,0,\sqrt{\frac{7}{12}},0,0,0,\sqrt{\frac{5}{12}}\cos 4\theta \right)^\top
\end{equation}

As done in the construction of coefficient vectors in the  $2D$ case, we can get rid of the angle $\theta$ by replacing it by a vector
$c=(c_0,c_1) = \left(\sqrt{\frac{5}{12}}\cos 4\theta,\sqrt{\frac{5}{12}}\sin 4\theta\right)$.
\begin{eqnarray}
a &=& \sqrt{\frac{7}{12}}(0,0,0,0,1,0,0,0,0)^\top \\
&+& c_0(0,0,0,0,0,0,0,0,1)^\top\\
&+& c_1(1,0,0,0,0,0,0,0,0)^\top
\end{eqnarray}
With this equation, all frames having a vector equal to $z$ can be represented by the $2D$ vector $c$. As in the $2D$ case it comes with a norm constraint: $c_0^2+c_1^2=\frac{5}{12}$.

\begin{figure}
 \centerline{
 \includegraphics[width=8cm]{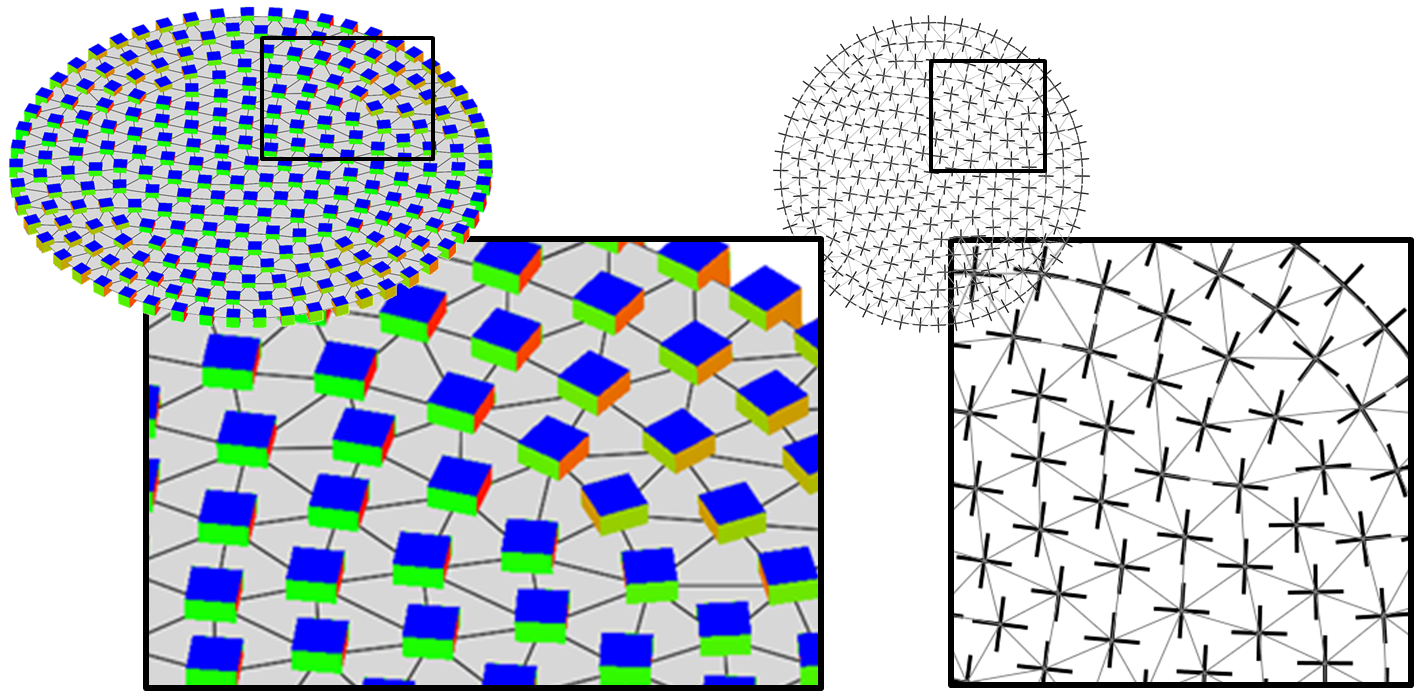}
 }
\caption{A $3D$ frame field produced on a ($2D$) disk (Left) produces the $2D$ frame field we could obtain from the $2D$ algorithm (Right).}
  \label{fig:flatdisk}
\end{figure}

The variable $c$ defines the rotation of the frame around the surface normal i.e. a $2D$ frame field. The optimization of this $2D$ frame field using $c$ as variables is exactly what we did in $2D$ by introducing the coefficient vector $a$. Our $3D$ solution restricted to the object boundary is therefore almost \footnote{The boundary has curvature that was not assumed in our $2D$ frame fields.} equivalent to our $2D$ solution (Figure~\ref{fig:flatdisk}).



{\bf Case 2:} the normal is not equal to the $z$ axis.

To handle this more general case, we rotate the constraint. If we want the vector $\vec{n}$ to be preserved, we first compute a rotation that brings $z$ axis to $\vec{n}$.
From this rotation, we compute the corresponding $9D$ rotation matrix $R_B$, and derive the constraints:
\begin{eqnarray}
\label{eq:3Dboundaryconstraint}
a &=& \sqrt{\frac{7}{12}}R_B(0,0,0,0,1,0,0,0,0)^\top \\
&+& c_0R_B(0,0,0,0,0,0,0,0,1)^\top \\
&+& c_1R_B(1,0,0,0,0,0,0,0,0)^\top
\end{eqnarray}

This expression of the normal constraint gives us a set of $9$ linear equations per boundary vertex.
It introduces two new variables $c_0$ and $c_1$, and a quadratic constraint $c^\top c=5/12$.

{\bf Note} As in the $2D$ case, the boundary constraint has a simpler expression \cite{Huang:2011} : $a^\top  R_B (0,0,0,0,1,0,0,0,0)^\top=\sqrt{\frac{7}{12}}$ that is valid only if all $a^i$ are feasible. Consequently, it cannot be used safely during the initialization step (see Figure~\ref{fig:boundarytet}).

\subsubsection*{Non smooth vertices} Frames of vertices located on hard edges have to conform to more than one normal. These vertices have multiple normal constraints, we pick two normals that are almost orthogonal, perturb them (by rotations around their cross product vector) to make them orthogonal, and compute the rotation that brings $x$ to the first normal, and $y$ to the second normal. We compute the corresponding coefficient space rotation $R_B$ and fix the frame coefficient vector $a^i$ to $R_B\tilde{a}$.

\subsection{Implementation}
\label{sec:implementation3D}

We have to minimize our objective function (eq.~\eqref{eq:smoothNRJ3D}) with linear equality constraints on boundary vertices eq.~\eqref{eq:3Dboundaryconstraint}, quadratic equality constraints $c^i\cdot {c^i}^\top =1$ on boundary vertices, and the constraint that each $a^i$ is feasible.

As in the $2D$ case (in \S\ref{sec:implementation2D}), our minimization algorithm (Algo.~\ref{alg:overview}) is formulated as a series of least squares problems (minimize $\|AX-b\|^2$), where $A$ and $b$ are constructed without the feasibility constraint of $a^i$ at the first iteration (initialization), and with a linear approximation of it in the subsequent iterations (smoothing iterations).

\begin{itemize}
\item {\bf Initialization:} 
Our variable vector $X$ must represent the representation vectors $a^i$ but also the $c^i$ variables introduced to express the boundary constraint. To do so, we first reorder vertices to have boundary vertices first \footnote{It is possible to increase the performances by $\approx 30\%$ by doing a Hilbert sort in the boundary vertices block, and another one in the free vertices block}. We can then organize the variable vector $X$ by blocks: $X[9i+d] = a^i_d$, and $X[9n_v+2i+d] = c^i_d$ where $n_v$ is the number of vertices. 

As in the $2D$ case, the matrix $A$ and the vector $b$ are constructed iteratively by adding new equations for the objective function (algorithm~\ref{alg:smoothing_terms}) and the boundary constraints (algorithm~\ref{alg:normal_constraints}). In our approach, we do not explicitly enforce the feasibility of $c^i$ ($c^i\cdot {c^i}^\top=5/7$), but it will be indirectly respected by the feasibility of $a^i$.

The projection of $a^i$ on the set of feasible coefficient vectors is no longer a simple normalization. 
Instead we perform, for each vertex, a gradient descent (algorithm~\ref{alg:closestframe}) initialized by $\tilde{a}$. 
More precisely, starting with $\tilde{a}$ we rotate our current frame gradually in order to minimize the distance between the current frame function and the function to be projected.
The gradient is evaluated by calculating the variation of the $L^2$ norm induced by infinitesimal rotation matrices with Euler's angles.

\item {\bf Smoothing iteration:} 
For the linearized feasibility constraint of $a^i$, we must also add $3$ extra variables per vertex to our system. These variables account for the position in a local basis of the tangent space of the $3D$ manifold of feasible $a^i$.
The introduction of these constraints in the matrix $A$ is detailed in algorithm~\ref{alg:local_opt}.
\end{itemize}

This frame field design algorithm can be implemented without being expert in spherical harmonics.
We give explicit construction of matrices $R_B, E^x, E^y, E^z$ in the appendix. The system to solve $A^\top AX=A^\top b$ is simply a linear system with a positive definite matrix. 
We use the OpenNL library \cite{OpenNL} because the system can be directly constructed from the equations (lines of $A$ and elements of $b$).

In order to keep the algorithm easy to read, we did not detail how to lock frames for vertices with multiple normal constraints.


\begin{algorithm}[!h]
\LinesNumbered
\SetAlgoLined
\KwIn{\begin{itemize}\item A tetrahedral mesh $\mathcal{M}$ with:\begin{itemize}\item $n_v$ vertices including $n_l$ vertices with normal constraint \item edges $\mathcal{E}$ \end{itemize} \item number of smoothing iterations $N$\end{itemize}}
\KwOut{A frame $f^i$ for each vertex $i$}

\vspace{3mm}
sort($\mathcal{M},\mathcal{E}$); // {\it vertex $i$ is a boundary vertex $\iff i<n_l$}\\
\vspace{1mm}
\ForEach{$I\in 0\dots N$}{
    // {\it $A$ and $b$ will be constructed iteratively} \\
    create matrix $A$ with $0$ rows and $9n_v+2n_l+3n_v$ columns\;
    create vector $b$ of size $0$\;
    \vspace{1mm}
    add\_smoothing\_terms($\mathcal{M}$, $A$, $b$)\;
    add\_normal\_constraints($\mathcal{M}$, $A$, $b$)\;
	// {\it add constraints only if we are in a smoothing iteration}\\    
    \If{$I>0$} {
        add\_local\_optim\_constraints($\{a^i\}$, $\mathcal{M}$, $A$, $b$)\;
    }
    \vspace{1mm}
    // {\it solve $A^\top AX=A^\top b$}\\
    $X\leftarrow$ call\_least\_squares\_solver($A,b$)\; 
    \vspace{1mm}
    // {\it find the frame for each vertex}\\
    \ForEach{$i<n_v$}{
        $a^i\leftarrow X[9i\dots 9i+8]$\;
        $(f^i, a^i) \leftarrow$ closest\_frame($a^i$)\;
    }
}
\caption{Frame field optimization}
\label{alg:overview}
\end{algorithm}

\begin{algorithm}[!h]
\LinesNumbered
\SetAlgoLined
\KwIn{A tetrahedral mesh $\mathcal{M}$, matrix $A$, row $b$}
\KwOut{Modified matrix $A$ and vector $b$}
// {\it enforcing normal constraints by quadratic penalty} \\
\ForEach{$i<n_l$} {
    estimate normal $n$ at vertex $i$\;
    find Euler angles $(\alpha,\beta,\gamma)$ to rotate $z$-axis to $n$\;
    find $9\times9$ rotation matrix $R_B$; // \textit{see appendix~\ref{sec:9drotation}}\\
    $h_0\leftarrow R_B\times(1,0,0,0,0,0,0,0,0)^\top$\;
    $h_4\leftarrow R_B\times(0,0,0,0,1,0,0,0,0)^\top$\;
    $h_8\leftarrow R_B\times(0,0,0,0,0,0,0,0,1)^\top$\;
    $\lambda\leftarrow 100$; // \textit{quadratic penalty multiplier} \\
    \ForEach{$d \in 0\dots 8$}{
        create vector $row$\;
        $row[9i+d]\leftarrow \lambda$\;
        $row[9n_v+2i+0]\leftarrow \lambda h_0[d]$\;
        $row[9n_v+2i+1]\leftarrow \lambda h_8[d]$\;
        $A.$add\_row($row$)\;
        $b.$push($\lambda \sqrt{7/12}h_4[d]$)\;
    }
}
\caption{add\_normal\_constraints}
\label{alg:normal_constraints}
\end{algorithm}

\begin{algorithm}[!h]
\LinesNumbered
\SetAlgoLined
\KwIn{A tetrahedral mesh $\mathcal{M}$, matrix $A$, row $b$}
\KwOut{Modified matrix $A$ and vector $b$}
\ForEach{$ij \in \mathcal{E}$}{
    \ForEach{$d \in 0\dots 8$}{
        create vector $row$\;
        $row[9i+d]\leftarrow 1$\;
        $row[9j+d]\leftarrow -1$\;
        $A.$add\_row($row$)\;
        $b.$push(0)\;
    }
}
\caption{add\_smoothing\_terms}
\label{alg:smoothing_terms}
\end{algorithm}

\begin{algorithm}[!h]
\LinesNumbered
\SetAlgoLined
\KwIn{Previous solution $\{a^i\}_{i<n_v}$, tetrahedral mesh $\mathcal{M}$, matrix $A$, row $b$}
\KwOut{Modified matrix $A$ and vector $b$}
\ForEach{$i<n_v$}{
    $c_x \leftarrow E^x\times a^i$; // \textit{see appendix~\ref{sec:linearization}}\\
    $c_y \leftarrow E^y\times a^i$\;
    $c_z \leftarrow E^z\times a^i$\;
    $\lambda\leftarrow 100$; // \textit{quadratic penalty multiplier} \\
    \ForEach{$d\in 0\dots 8$}{
        create vector $row$\;
        $row[9 i + d] \leftarrow \lambda$\;
        $row[9n_v+2n_l+3i+0] \leftarrow -\lambda c_x[d]$\;
        $row[9n_v+2n_l+3i+1] \leftarrow -\lambda c_y[d]$\;
        $row[9n_v+2n_l+3i+2] \leftarrow -\lambda c_z[d]$\;
        $A.$add\_row($row$)\;
        $b.$push($\lambda a^i[d]$)\;
    }
}
\caption{add\_local\_optim\_constraints}
\label{alg:local_opt}
\end{algorithm}

\begin{algorithm}[!h]
\LinesNumbered
\SetAlgoLined
\KwIn{9-component vector $q$}
\KwOut{A frame $f$ and its representation vector $a$}

$f\leftarrow\tilde{f}$\;
$a\leftarrow\tilde{a}$\;
$s\leftarrow 10^{-1}$; // \textit{optimization step size}\\
$\varepsilon \leftarrow 10^{-4}$; // \textit{step threshold}\\
$q\leftarrow q/|q|$\;
\While{True} {
    $g\leftarrow (q^\top E^x_B a, q^\top E^y_B a, q^\top E^z_B a)$; // \textit{gradient in point $a$}\\
    \If{$\|g\|<\varepsilon$} {
        break;
    }
    $R_B\leftarrow R_B^x(s\cdot g[0]) \times R_B^y(s\cdot g[1]) \times R_B^z(s\cdot g[2])$\;
    $R\leftarrow R^x(s\cdot g[0]) \times R^y(s\cdot g[1]) \times R^z(s\cdot g[2])$\;
    $a\leftarrow R_B a$\;
    $f\leftarrow R f$\;
}
 return $f, a$\;

\caption{closest\_frame}
\label{alg:closestframe}
\end{algorithm}

\subsection{Results}
\label{sec:results}

It is impossible to compare frame field design algorithms only from the images and results presented in the state of the art papers.
First of all, our implementation of~\cite{Huang:2011} has significantly better performances compared to what was presented in the original paper.
Next, Li \emph{et al.}~\cite{Li:2012} did not present any frame field results, but only hex meshes that was the main focus of their work.
Therefore, we implemented both methods; there are few points worth noting:


\emph{Sampling:} In previous works the frame fields were sampled either on each tet face or on each tet. Instead we sample it on vertices, otherwise we would not be able to compare corresponding energies.

\emph{Gimbal Lock:} Both Huang and Li use Euler angles as variables in their L-BFGS optimization, which have numerical issues when the angles are close to the gimbal lock.
Note that each frame can be represented by 48 triplets of equivalent Euler angles. In our implementation we choose the triplet maximizing the distance to the nearest gimbal lock.

\emph{Rendering:} For rendering purposes, we rely on a combination of techniques (Figure~\ref{fig:rendering}) to show the spherical harmonics field, the frame field (locally and globally) and the field topology.

\subsubsection{Comparison with Huang's method}
\label{sec:ch1eval}

\begin{figure}
 \centerline{\includegraphics[width=8cm]{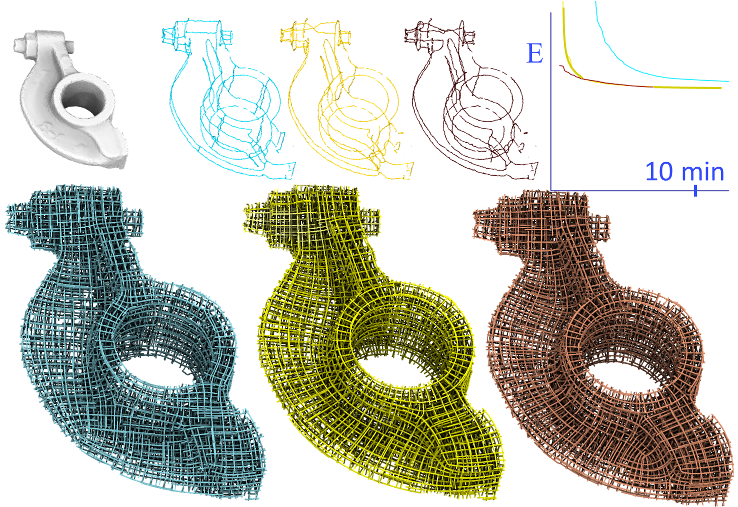}}
 \centerline{\includegraphics[width=8cm]{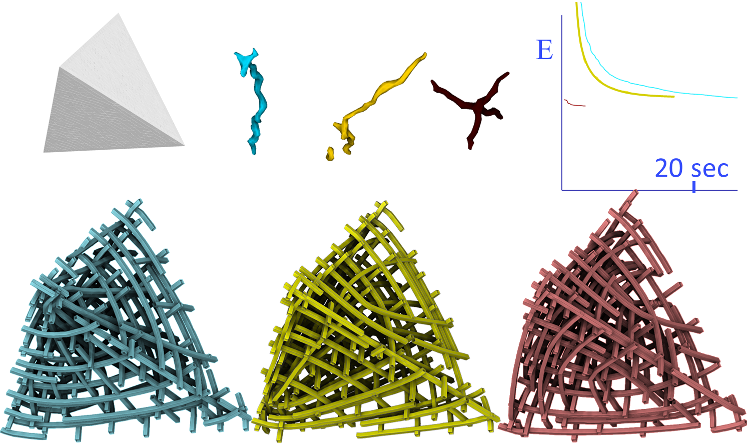}}
\caption{Comparison of initialization of Huang's algorithm with their weight of boundary penalty term $(10^{-2})$ (left/blue), with a much higher weight $(10^2)$ (middle/yellow), and our initialization (right/red). All smoothing iterations are performed by Huang's algorithm. For the rocker model, the energies (we take the run with our initialization for $100\%$) are respectively $94,7\%$, $101,6\%$ and (obviously) $100\%$. For the tet model, we obtain $91\%$, $89,7\%$ and $100\%$.}
  \label{fig:huanginitW}
\end{figure}

Recall that Huang \emph{et al.} proposed a method in two steps:
\begin{itemize}
\item find an initial frame field by solving a linear system and projecting the solution onto the manifold of feasible solutions
\item represent each frame by a triplet of Euler angles and optimize the smoothness using an L-BFGS descent method.
\end{itemize}
Our implementation produces results very similar to those presented in~\cite{Huang:2011}, but with significantly better timings.
For example, the rockarm (Figure~\ref{fig:huanginitW}) with one million tetrahedra takes about 10 minutes on a single thread application on a Dell M6600 laptop compared
to 155 minutes obtained by Huang \emph{et al.} on a two-thread i7 processor.

Huang's initialization is very similar to ours, however their boundary condition is not sufficient in this case (it requires the SH coefficients to be feasible).
Moreover, they enforce the boundary condition with a penalty term that is very light ($10^{-2}$ weight).
As a consequence, their initial fields are almost constant everywhere (it maximizes the smoothness), with a topology very far from being optimal.
The smoothing iterations are performed with much higher weight ($10^3$) of the penalty term using L-BFGS.


After the initialization step we measured the deviation of the field from the given constraints on the rockarm model.
Note that the penalty term being the sum of deviations over all vertices, we can conclude that deviation at a given vertex belongs to $[0, 2\sqrt{7/12}]$.
Thus on the rockarm the initial frame field has the average deviation of $0.34$, whereas the maximum frame deviation is $0.96$.

If we use a much higher penalty weight to enforce the boundary constraint, we obtain an initial frame field with average deviation from constraints
equal to $0.07$ and maximum deviation equal to $0.75$. The field has better topology and L-BFGS converges faster for this initialization.
The initialization provided by our method has average deviation from constraints equal to $10^{-8}$ with maximum deviation of $10^{-7}$.

Figure~\ref{fig:huanginitW} gives an illustration, it compares three methods: Huang's algorithm (left image) Huang's algorithm with much higher penalty weight (middle image) and
our initialization followed by Huang's smoothing iterations (right).


Huang's paper is the pioneer work and the main contribution in~\cite{Huang:2011} was the introduction of the energy used in frame field optimization,
however the initialization is not very good and smoothing stage was also outperformed by later works (see \S\ref{sec:smoothcmp}).

\subsubsection{Comparison with Li's method}
\label{sec:ch2eval}

Li's initialization computes a $2D$ frame field on the surface, then propagates it inside the volume by advancing front. As a consequence, the initial field perfectly matches boundary constraints, but is discontinuous across its medial axis.

The smoothing iterations are performed by L-BFGS. It acts on a new set of variables that characterizes, per vertex, the rotation that brings the reference frame to the current frame.
For the frames located inside the object, variables are the Euler angles as in Huang's method. For frames located on the object boundary, the rotation is characterized by a single rotation angle around the normal vector.

The frame field results presented in their paper were limited to an ellipsoid and a sphere (frame field design was not the primary objective).
We guess that most of presented results were not fully automatically generated frame fields, as they wrote: {\it ``For instance, we use guiding boxes to modify the frames inside the narrow ears of Bunny (Figure 13-a) and the head of rock arm (Figure 13-b) to reduce singularities''}.

Moreover, before implementing the method, we thought that their algorithm was strongly limited by the original field topology from the sentence: {\it ``However, our propagation-based frame field initialization likely generates singular edges around the medial axis of the volume, and most of them cannot be eliminated by frame optimization.''}
Surprisingly, our implementation of their method is able to generate smooth frame fields automatically in most situations, even when the topology of the initial field is complex close to the medial axis. 
Our implementation is slightly different:
\begin{itemize}
\item we initialize the $2D$ field by our $3D$ algorithm restricted to boundary vertices
\item we sample the field on vertices
\item and we prevent gimbal locks by a proper initialization of Euler angles.
\end{itemize}

\begin{figure}
 \centerline{\includegraphics[width=8cm]{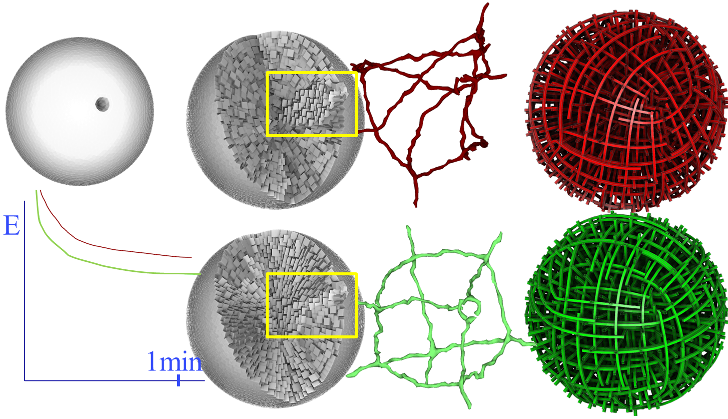}}

\caption{Li's algorithm (red) is compared to our algorithm (green) on a one-finger bowling ball. The hole has a huge impact on the initialization due to the advancing front approach (second column, inside the yellow box). As a result, their initialization provides a field with a poor topology and smoothing iterations are not sufficient to find the expected topology (like ours). Our final energy is $86\%$ of theirs.}
  \label{fig:LiinitW}
\end{figure}

The only real failure case we found using their method is due to the front propagation algorithm: when a boundary frame is copied to a large number on inner samples.
In this case, the L-BFGS solver is sometimes locked with a bad topology (see Fig.~\ref{fig:LiinitW}).

\begin{figure}
 \centerline{\includegraphics[width=8cm]{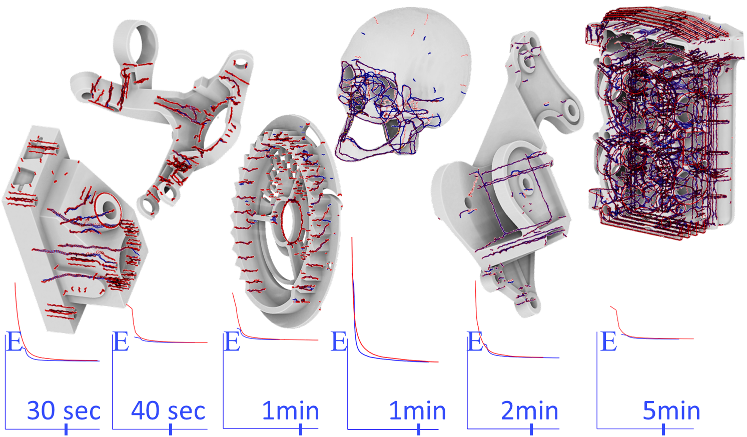}}
\caption{Comparison of two fields generated by Li \emph{et al.} smoothing iterations: using their initialization (red) or ours (blue). Our energy in percent of theirs is $99\%$, $99.87\%$, $100.5\%$, $99.97\%$, $99\%$, $99.6\%$. The difference is always very low ($<1\%$) but always in our advantage except for the third model.}
  \label{fig:cmpli}
\end{figure}

\begin{figure}
 \centerline{\includegraphics[width=8cm]{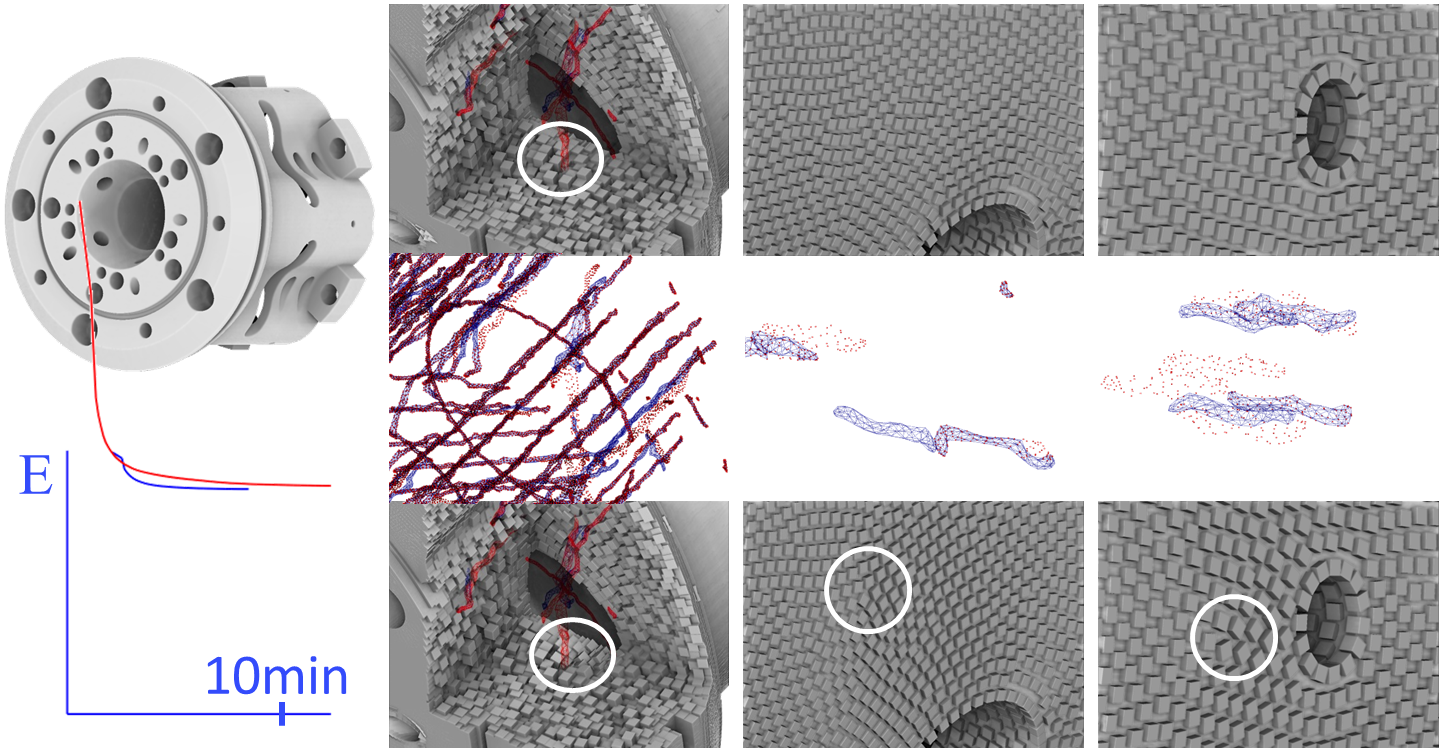}}
\caption{Comparison of two fields generated by Li's smoothing iterations: using their initialization (red) or ours (blue). Our energy in percent of theirs is $98.5\%$. Close-ups show the field where the singularity graphs diverge: one is inside the volume (leftmost) and others are on the object boundary. Our results are on the top row and theirs are on the bottom row with singularity encircled in white. }
  \label{fig:cmpli2}
\end{figure}

On more complex examples, we have compared their algorithm against our initialization followed by their smoothing iterations.
In most cases, we obtain an energy that is a bit better (Fig~\ref{fig:cmpli}). We also observed that their topology often differs from ours (Fig~\ref{fig:cmpli2}), so we conjecture that our topology is somehow better. However, the quality of the field topology depends on the application, and is not well evaluated by the energy, even for topologies very far from being optimal (see e.g. Fig~\ref{fig:huanginitW}).

\subsubsection{Comparison of smoothing iterations}
\label{sec:smoothcmp}

In the previous sections we have shown (Fig.~\ref{fig:huanginitW} and \ref{fig:LiinitW}) that our method provides the best initialization,
however our smoothing iterations are not clearly better than others.

\begin{figure}
 \centerline{\includegraphics[width=8cm]{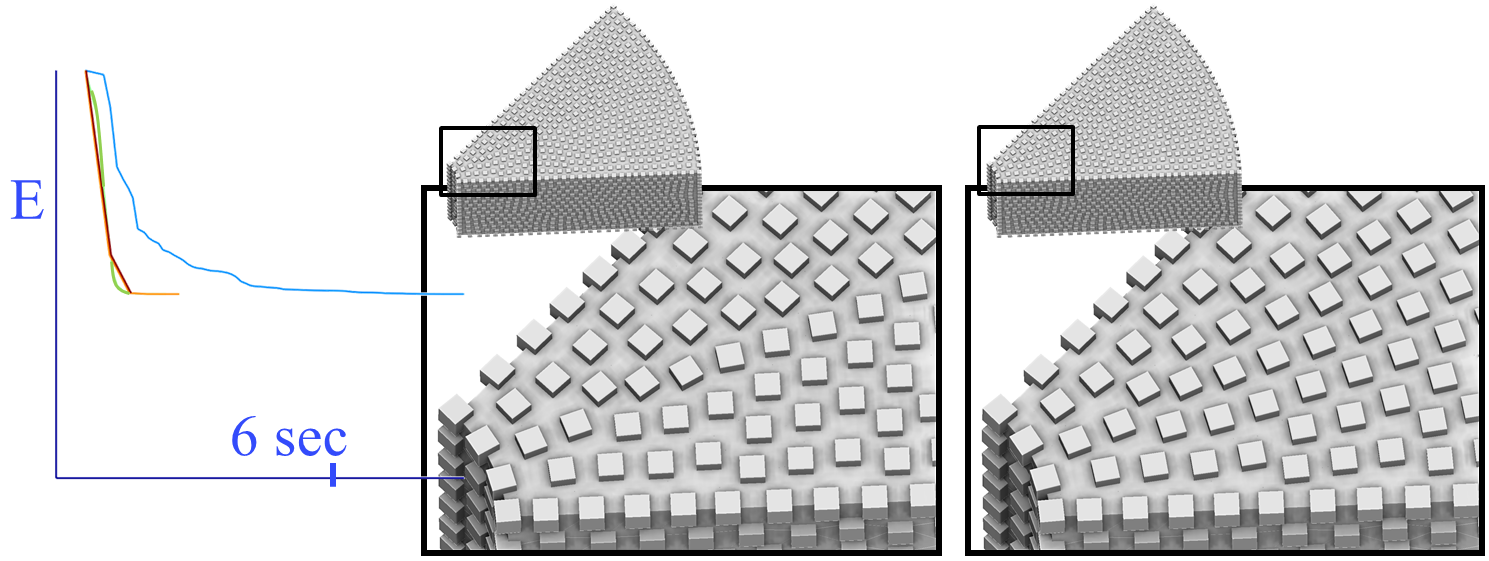}}
 \centerline{\includegraphics[width=8cm]{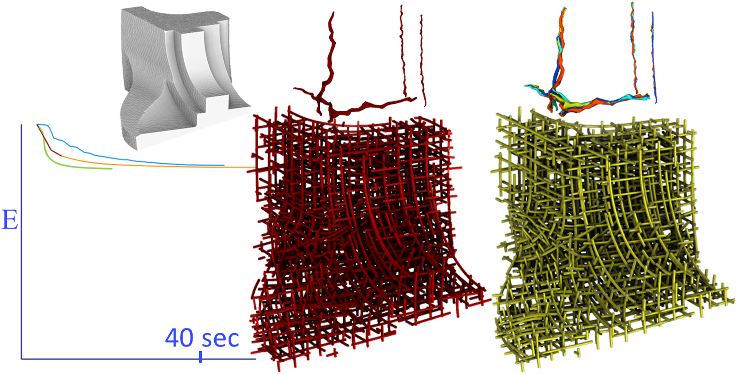}}
 \centerline{\includegraphics[width=8cm]{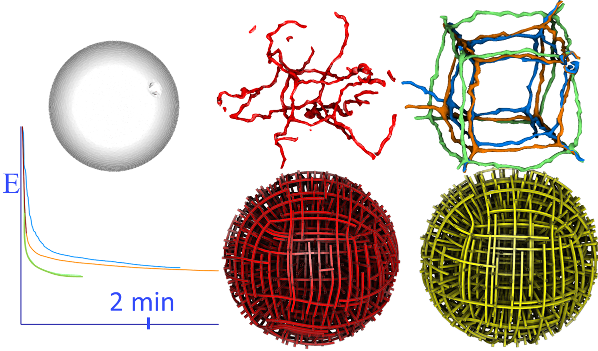}}
\caption{Comparison of smoothing iteration algorithms combined with our initialization algorithm. We compare Huang's method (blue), Li's method (green), our method (orange), and our method limited to two iterations (red).
Top row compares the initialization (left) with the field after two iterations of our algorithm (right).
Middle and bottom rows compare the field with two iterations of our smoothing algorithm with other smoothing strategies. Singularity graphs reflect nicely the convergence of thee algorithms. 
We obtain energy (we take Li's result for $100\%$) of resp. $99.98\%$, $100\%$, $99.98\%$ and $100.5\%$ on the sector, $101.5\%$, $100\%$, $100.4\%$, and $106.5\%$ on the fandisk, and $116\%$, $100\%$, $109\%$, $185\%$ on the one-finger bowling ball.}
  \label{fig:smoothiter}
\end{figure}

Figure~\ref{fig:smoothiter} shows a comparison of three different smoothing strategies (our linearization, L-BFGS proposed by Huang \emph{et al.} and L-BFGS proposed by Li \emph{et al.}).
In all three cases we use our method to initialize the field.
L-BFGS smoothing proposed by Huang \emph{et al.} is the slowest in all test cases.
First of all, in our implementation the time to evaluate the energy and the gradient is four times slower for the method by Huang \emph{et al.} than for the method by Li \emph{et al.}
Second, the cruicial difference between these two methods is the way to enforce the boundary alignment: Huang \emph{et al.} use a penalty term, whereas Li \emph{et al.}
use the set of variables directly satisfying the boundary constraints.
In our test we noticed that usage of penalty terms increases the number of iterations to converge and interferes with topological choices to be made, leading to inferior final fields.

We also noticed that the behaviour of linearization changes in function of how far the initialization is from the final solution.
\begin{itemize}
\item First row of figure~\ref{fig:smoothiter} shows a simple case without topology changes, the smoothing iterations change the field geometry only.
In this case the linearization of feasibility constraints works flawlessly, in two iterations the method converges, taking about the same time as the smoothing by Li \emph{et al.}
\item Middle row shows a second case, where few topology changes must be made. It slowes the linearization down, even if two iterations produce a reasonably good field.
\item Finally, the bottom row shows the case where the initial topology is really bad.
The linearization method fails on this model: two first iterations are still very far from the final solution and to reach the minimum it requires four times more time than the method by Li \emph{et al.}
\end{itemize}

We conclude that the best solution is our initialization followed by the optimization of Li \emph{et al.}.
In practice, the implementation of our linearization smoothing iterations is almost free (incremental with respect to our initialization algorithm),
whereas Li \emph{et al.} smoothing algorithm is more difficult to implement.
Moreover, in most cases few iterations of linearization steps suffice to obtain a fairly good field.
As a consequence, we suggest starting with our smoothing iterations (almost free to implement), then possibly replace it with Li \emph{et al.} smoothing algorithm if performances are not sufficient.

\begin{figure*}[!p]
\begin{minipage}[!t]{.46\linewidth}
\centerline{\includegraphics[width=\linewidth]{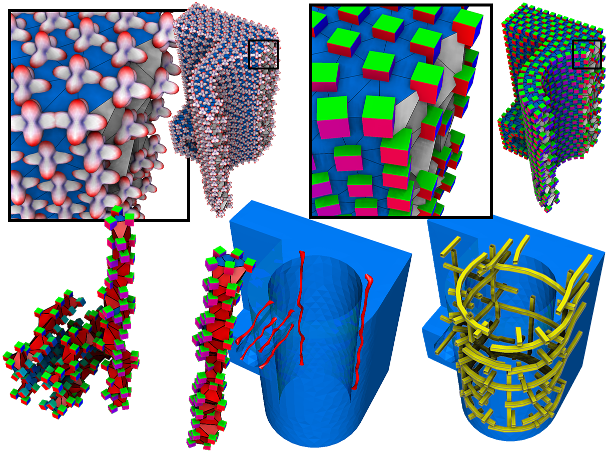}}
\caption{Our results are shown using combinations of the following rendering techniques. We can plot for each vertex $i$ its $F^i$ (upper left), or its frame as a cube(upper right). We can show the singular tets (lower left), or a smoothed and refined version (lower middle) to better see it in $3D$ (thanks to the lighting). The field inside the volume can be rendered by curved french fries.}
\label{fig:rendering}
\end{minipage}
\hspace{.05\linewidth}
\begin{minipage}[!t]{.46\linewidth}
\centerline{\includegraphics[width=8cm]{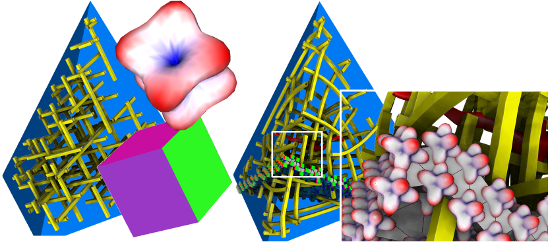}}
\caption{The initialization of \protect\cite{Huang:2011} (left) is a constant frame field whereas ours (right) is aligned to the boundary. Their $F^i$ are all equal, and very far from being feasible, making it possible to violate the boundary condition.}
\label{fig:boundarytet}
\end{minipage}

\vspace{5mm}


\centerline{ \includegraphics[width=\linewidth]{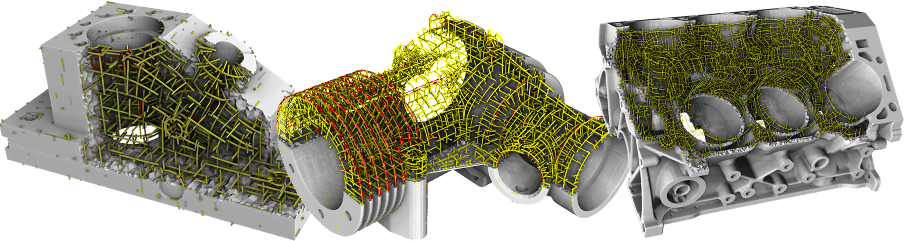}}
\caption{Results on CAD objects. Names are (from left to right): Anc, Crank, 40head.}
  \label{fig:result}

\vspace{5mm}

\centerline{\includegraphics[width=\linewidth]{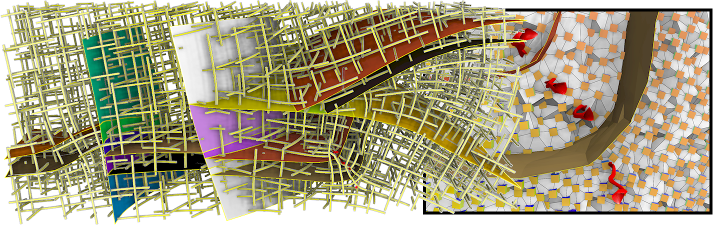}}
\caption{3D frame field constrained by faults in geological data.}
\label{fig:botela}

\end{figure*}

\section*{Future works}

\begin{figure}[!htbp]
\centerline{\includegraphics[width=\linewidth]{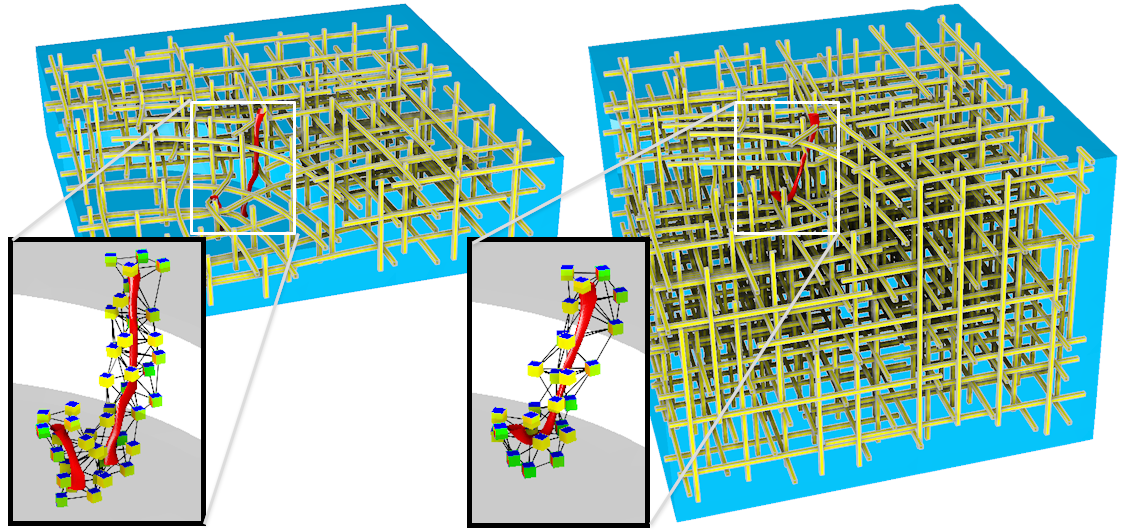}}
\caption{The thin object (left) frame field is suitable to produce a hex mesh: its singularity graph is basically two singularities of index $1/4$ and $-1/4$ extruded in the $z$ direction. The frame field of fat object (right) has a singularity graph that does not correspond to a hex mesh.}
  \label{fig:badsing}
\end{figure}

From an application point of view, our method makes it fast and easy to produce smooth frame fields. The smoothness is not necessary the optimal objective for applications such as hex remeshing (see Fig.~\ref{fig:badsing}), but can be used as a regulation term for more complex energies. It is also very easy to modify our method to add constraints inside a volume. Figure~\ref{fig:botela} shows a frame field constrained by faults in geological data. Note that the field is not constrained by the boundary of the model. Such a frame field is useful to steer element placement for fluid simulation used in oil exploration.

From a theoritical point of view, it would be interesting to better understand the shape of the feasible set of $a^i$ (the $3D$ manifold embedded in $9D$). It could help to find a better projection algorithm than our current gradient descent.

\section*{Conclusion}

This work unifies the frame field design problem in $2D$ and $3D$.
Both problems are formulated with a similar representation of frames, constraints and objective function. As a consequence, they can also be solved by similar algorithms.

From this analysis, we discovered that the best actual solution to produce smooth $3D$ frame fields is to initialize it by our proposed extension of \cite{kowalski2012}, followed by smoothing iterations of \cite{Li:2012}.
The main drawback of this solution is requirement to implement two very different approaches (a sparse linear system solver and a L-BFGS descent).
A fair alternative is to use our extension of \cite{kowalski2012}, it is simple to implement and requires a linear system solver only.
In practice for our models we perform only two or three linearization iterations, however if the initialization is a bad guess (e.g. the sphere), it can be insufficient.
With this approach we are able to generate (on a laptop) fields on the models up to few millions tetrahedra in less than 10 minutes, refer to Figure~\ref{fig:result} for an illustration.

\appendix

\newcommand{\atan}{\mathop{\mathrm{atan2}}}

\section{SH cookbook}

\subsection{$9D$ rotation}
\label{sec:9drotation}
Let us denote by $R^x$, $R^y$ and $R^z$ $3\times 3$ matrices of rotation around axes $x$, $y$ and $z$ respectively.
Any frame $f$ can be obtained as a composed rotation of the reference frame $\tilde{f}$, where the reference frame is the set of 6 unit vectors aligned with coordinate axes:
$$f = R^x(\alpha)\times R^y(\beta) \times R^z(\gamma) \times \tilde{f}.$$

If $(\alpha, \beta, \gamma)$ are Euler angles of rotation between a frame $f$ and $\tilde{f}$,
the representation vector $a$ is calculated as $a = R_B^x(\alpha)\times R_B^y(\beta) \times R_B^z(\gamma) \times \tilde{a}$,
where $R_B^x$, $R_B^y$ and $R_B^z$ are $9\times 9$ matrices of rotation defined as follows:
\begingroup\BigColSep
\begin{multline*}
R_B^z(\gamma)=\\
\begin{tiny}
\begin{bmatrix}
\cos(4\gamma)&    0&             0&             0&             0&             0&             0&             0&             \sin(4\gamma) \\
0&            \cos(3\gamma)&     0&             0&             0&             0&             0&             \sin(3\gamma)&     0         \\
0&            0&             \cos(2\gamma)&     0&             0&             0&             \sin(2\gamma)&     0&             0         \\
0&            0&             0&             \cos(\gamma)&        0&             \sin(\gamma)&        0&             0&             0         \\
0&            0&             0&             0&             1&             0&             0&             0&             0         \\
0&            0&             0&             -\sin(\gamma)&       0&             \cos(\gamma)&        0&             0&             0         \\
0&            0&             -\sin(2\gamma)&    0&             0&             0&             \cos(2\gamma)&     0&             0         \\
0&            -\sin(3\gamma)&    0&             0&             0&             0&             0&             \cos(3\gamma)&     0         \\
-\sin(4\gamma)&   0&             0&             0&             0&             0&             0&             0&             \cos(4\gamma)
\end{bmatrix}
\end{tiny}
\end{multline*}

\begin{multline*}
R_B^x(\pi/2) = \\
\begin{tiny}
\begin{bmatrix}
0&                  0&                0&                  0&                0&                 \sqrt{14}/4& 0&                 -\sqrt{2}/4& 0                \\
0&                  -3/4&           0&                  \sqrt{7}/4& 0&                 0&                 0&                 0&                 0                \\
0&                  0&                0&                  0&                0&                 \sqrt{2}/4&  0&                 \sqrt{14}/4& 0                \\
0&                  \sqrt{7}/4& 0&                  3/4&            0&                 0&                 0&                 0&                 0                \\
0&                  0&                0&                  0&                3/8&             0&                 \sqrt{5}/4&  0&                 \sqrt{35}/8\\
-\sqrt{14}/4& 0&                -\sqrt{2}/4&  0&                0&                 0&                 0&                 0&                 0                \\
0&                  0&                0&                  0&                \sqrt{5}/4&  0&                 1/2&             0&                 -\sqrt{7}/4\\
\sqrt{2}/4&   0&                -\sqrt{14}/4& 0&                0&                 0&                 0&                 0&                 0                \\
0&                  0&                0&                  0&                \sqrt{35}/8& 0&                 -\sqrt{7}/4& 0&                 1/8
\end{bmatrix}
\end{tiny}
\end{multline*}
\endgroup

$$
R_B^y(\beta) = R_B^x(\pi/2)\times R_B^z(\beta)\times R_B^x(\pi/2)^{\top}
$$

$$
R_B^x(\alpha) = R_B^y(\pi/2)^{\top} \times R_B^z(\alpha) \times R_B^y(\pi/2)
$$

These matrices are called Wigner D-matrices and the literature on their construction is vast~\cite{Collado1989323,Blanco199719,gna480229,jp953350u}.
However, as we are using degree 4 harmonics only, we performed the symbolic computation as follows.

The matrix $R^z_B(\gamma)$ is easy to compute, given a spherical harmonic $Y_{4,m}(\theta, \phi)$ we can rotate it around the $z$-axis by the angle $\gamma$
by evaluating $Y_{4,m}(\theta, \phi+\gamma)$.
Then the element $(i,j)$ of the matrix $R^z_B$ is simply $\int\limits_0^\pi\int\limits_0^{2\pi} Y_{4,i-5}(\theta,\phi)\cdot Y_{4,j-5}(\theta,\phi+\gamma) \sin\theta d\theta d\phi$.

Matrices $R^y_B(\beta)$ and $R^x_B(\alpha)$ are trickier to get, however we can use the fact that arbitrary rotation around the $y$-axis can be decomposed
into the rotation around $x$-axis by $\pi/2$ followed by a rotation around $z$-axis.
To rotate a spherical harmonics $Y_{4,m}(\theta, \phi)$ around the $x$-axis by $\pi/2$ we can perform the following substitution:
$$
(\theta, \phi) \leftarrow (\arccos(-\sin(\theta)\sin(\phi)), \atan(\cos(\theta),\sin(\theta)\cos(\phi))).
$$
Then the matrix $R_B^x(\pi/2)$ is calculated by evaluating double integrals of all possible products between basis functions and rotated functions.


\subsection{Linearization}
\label{sec:linearization}

In order to linearize the constraints in our linear solver we define matrices $E^x_B, E^y_B$ and $E^z_B$ as follows:

\begingroup\BigColSep
$$
E^x_B =
\begin{tiny}
\begin{bmatrix}
           0 &          0&            0    &        0   &         0       &     0 &           0&           -\sqrt{2} &           0 \\
           0 &          0&            0    &        0   &         0       &     0 &-\sqrt {7/2} &          0   &        -\sqrt{2} \\
           0 &          0&            0    &        0   &         0       &  -3/\sqrt{2} &           0& -\sqrt{7/2} &            0 \\
           0 &          0&            0    &        0   &  -\sqrt{10}       &     0 &        -3/\sqrt{2}&            0 &           0 \\
           0 &          0&            0    &  \sqrt{10}   &         0       &     0 &           0&            0 &           0 \\
           0 &          0&          3/\sqrt{2}    &        0   &         0       &     0 &           0&            0 &           0 \\
           0 & \sqrt{7/2} &           0     &     3/\sqrt{2}    &        0        &    0  &          0 &           0  &          0 \\
           \sqrt{2} &          0&  \sqrt{7/2}      &      0     &       0         &   0   &         0  &          0   &         0 \\
           0 &          \sqrt{2}&            0    &        0   &         0       &     0 &           0&            0 &           0
\end{bmatrix}
\end{tiny}
$$

$$
E^y_B =
\begin{tiny}
\begin{bmatrix}
         0 &           \sqrt{2}&            0&            0&            0&            0&            0&            0&            0 \\
        -\sqrt{2} &           0&  \sqrt{7/2}  &          0  &          0  &          0  &          0  &          0  &          0   \\
         0 &-\sqrt{7/2}  &          0  &        3/\sqrt{2}  &          0  &          0  &          0  &          0  &          0   \\
         0 &           0&         -3/\sqrt{2}&            0&            0&            0&            0&            0&            0 \\
         0 &           0&            0&            0&            0&     -\sqrt{10}&            0&            0&            0 \\
         0 &           0&            0&            0&      \sqrt{10}&            0&         -3/\sqrt{2}&            0&            0 \\
         0 &           0&            0&            0&            0&          3/\sqrt{2}&            0& -\sqrt{7/2}  &          0  \\
         0 &           0&            0&            0&            0&            0&  \sqrt{7/2}  &          0  &         -\sqrt{2}   \\
         0 &           0&            0&            0&            0&            0&            0&            \sqrt{2}&            0
\end{bmatrix}
\end{tiny}
$$
\endgroup

$$
E^z_B =
\begin{tiny}
\begin{bmatrix}
 0&  0&  0&  0&  0&  0&  0&  0&  4\\
 0&  0&  0&  0&  0&  0&  0&  3&  0\\
 0&  0&  0&  0&  0&  0&  2&  0&  0\\
 0&  0&  0&  0&  0&  1&  0&  0&  0\\
 0&  0&  0&  0&  0&  0&  0&  0&  0\\
 0&  0&  0& -1&  0&  0&  0&  0&  0\\
 0&  0& -2&  0&  0&  0&  0&  0&  0\\
 0& -3&  0&  0&  0&  0&  0&  0&  0\\
-4&  0&  0&  0&  0&  0&  0&  0&  0
\end{bmatrix}
\end{tiny}
$$

It is easy to verify that these matrices are chosen to verify the following equations for small rotations $\alpha, \beta, \gamma$:
\begin{align*}
R^x_B(\alpha) &= I_{9\times 9}+\alpha E^x_B + o(|\alpha|)\\
R^y_B(\beta)  &= I_{9\times 9}+\beta E^y_B + o(|\beta|)\\
R^z_B(\gamma) &= I_{9\times 9}+\gamma E^z_B + o(|\gamma|).
\end{align*}

Finally, for small rotations the multiplication is commutative:
\begin{align*}
R_B(\alpha, \beta, \gamma) & = R^x_B(\alpha)\times R^y_B(\beta) \times R^z_B(\gamma) = \\
&= I_{9\times 9} + \alpha  E^x_B + \beta  E^y_B  + \gamma  E^z_B + o(|\alpha|+|\beta|+|\gamma|).
\end{align*}

\bibliographystyle{acmtog}
\bibliography{framefield}

\received{September 2008}{March 2009}

\end{document}